%% file: TMM.tex
\documentclass[lettersize,journal]{IEEEtran}
\usepackage{amsmath,amsfonts}
\usepackage[linesnumbered,ruled]{algorithm2e}
\usepackage{algorithmicx}
\usepackage{array}
\usepackage{textcomp}
\usepackage{stfloats}
\usepackage{url}
\usepackage{verbatim}
\usepackage{graphicx}
\usepackage{cite}
\usepackage{booktabs}
\usepackage{colortbl}
\usepackage{multirow}
\usepackage{arydshln}
\usepackage{bm}
\usepackage{xspace}
\usepackage{subcaption}
\usepackage[dvipsnames]{xcolor}
\usepackage{tikz}
\usetikzlibrary{shapes,snakes,backgrounds,patterns,arrows.meta}

\usepackage{array}
\usepackage{pifont}
\usepackage{comment}
\usepackage{mathtools}
\usepackage{diagbox}
\usepackage[pagebackref,breaklinks,colorlinks,citecolor=blue]{hyperref}
\usepackage{listings}
\usepackage{pgfplots, pgfplotstable}
\usepgfplotslibrary{groupplots}
\pgfplotsset{compat=newest}
\usepgfplotslibrary{fillbetween}

\SetCommentSty{mycommfont}

\SetKwInput{KwInput}{Input}                
\SetKwInput{KwOutput}{Output}              

\newcommand{\tabincell}[2]{\begin{tabular}{@{}#1@{}}#2\end{tabular}}
\newcommand{\Th}[1]{\textsc{#1}}
\newcommand{\mr}[2]{\multirow{#1}{*}{#2}}
\newcommand{\mc}[2]{\multicolumn{#1}{c}{#2}}
\newcommand{\tb}[1]{\textbf{#1}}

\newcolumntype{H}{>{\setbox0=\hbox\bgroup}c<{\egroup}@{}}

\usepackage{pgfplots, pgfplotstable}
\pgfplotsset{compat=newest}
\usepgfplotslibrary{fillbetween}

\newcommand{\oxf}{MidnightBlue}

\newcommand{\paris}{RedViolet}

\newcommand{\valc}{PineGreen}

\newcommand{\ourscolor}{Maroon}
\newcommand{\ourscolors}{NavyBlue}
\newcommand{\howcolor}{Black}
\newcommand{\csdcolor}{Black}

\newcommand{\oxfmark}{o}
\newcommand{\parmark}{triangle}
\newcommand{\valmark}{square}

\newcommand{\leg}[1]{\addlegendentry{#1}}

\tikzset{every mark/.append style={solid}}
\pgfplotsset{
	grid=both, width=\linewidth, try min ticks=5,
	legend cell align=left, 
	legend style={fill opacity=0.8},
	ylabel near ticks,
	xlabel near ticks,
	every tick label/.append style={font=\footnotesize},
}

\pgfplotsset{
	oxmed/.style={thick, color=\oxf, mark=o},
	oxhard/.style={thick, dashed, color=\oxf, mark=o},
	pamed/.style={thick, color=\paris, mark=star}, 
	pahard/.style={thick, dashed, color=\paris, mark=star},
	val/.style={thick, color=\valc, mark=x},
	oursoxf/.style={thick=20pt, color=\ourscolor, mark=o},
	oursoxfs/.style={thick=20pt, dashed, color=\ourscolors, mark=triangle},
	howoxf/.style={thick=10pt, dashed, color=\howcolor, mark=\oxfmark},
	csdoxf/.style={thick, dotted, color=\csdcolor, mark=\valmark},
	ourspar/.style={thick, color=\ourscolor, mark=\parmark},
	howpar/.style={thick, dashed, color=\howcolor, mark=\parmark},
	oursval/.style={thick, color=\ourscolor, mark=\valmark},
	howval/.style={thick, dashed, color=\howcolor, mark=\valmark},
	numean/.style={thick, color=\ourscolor, mark=none},
	numin/.style={thick, color=gray, mark=none},
	wid/.style={thick, color=\ourscolor, mark=o, mark size=0.5pt},
	woid/.style={thick, densely dashdotted, color=RedOrange, mark=o, mark size=0.5pt},
}

\makeatletter
\DeclareRobustCommand\onedot{\futurelet\@let@token\@onedot}
\def\@onedot{\ifx\@let@token.\else.\null\fi\xspace}

\def\eg{\emph{e.g}\onedot} 
\def\ie{\emph{i.e}\onedot}

\def\etal{\emph{et al}\onedot}

\def\adl@drawiv#1#2#3{%
	\hskip.5\tabcolsep
	\xleaders#3{#2.5\@tempdimb #1{1}#2.5\@tempdimb}%
	#2\z@ plus1fil minus1fil\relax
	\hskip.5\tabcolsep}
 
\newcommand{\cdashlinelr}[1]{%
	\noalign{\vskip0.7\aboverulesep
		\global\let\@dashdrawstore\adl@draw
		\global\let\adl@draw\adl@drawiv}
	\cdashline{#1}
	\noalign{\global\let\adl@draw\@dashdrawstore
		\vskip0.7\belowrulesep}}
  
\makeatother

\input{math_commands.tex}

\begin{document}

\title{Structure Similarity Preservation Learning for Asymmetric Image Retrieval}

\author{Hui Wu, Min Wang, Wengang Zhou,~\IEEEmembership{Senior Member,~IEEE}, Houqiang Li,~\IEEEmembership{Fellow,~IEEE}
\thanks{This work is supported by National Natural Science Foundation of China under Contract 62102128 and 62021001. It was also supported by GPU cluster built by MCC Lab of Information Science and Technology Insti-tution, USTC, and the Supercomputing Center of the USTC.}
\thanks{Hui Wu, Wengang Zhou, and Houqiang Li are with the CAS Key Laboratory of Technology in Geospatial Information Processing and Application System, Department of Electronic Engineering and Information Science, University of Science and Technology of China, Hefei, 230027, China (e-mail: wh241300@mail.ustc.edu.cn; zhwg@ustc.edu.cn; lihq@ustc.edu.cn). Min Wang is with the Institute of Artificial Intelligence, Hefei Comprehensive National Science Center, Hefei 230030, China (e-mail: wangmin@iai.ustc.edu.cn).}
\thanks{Corresponding authors: Min Wang and Wengang Zhou.}}

\markboth{IEEE Transactions on Multimedia,~Vol.~**, No.~**, October~2023}%
{Shell \MakeLowercase{\textit{et al.}}: A Sample Article Using IEEEtran.cls for IEEE Journals}

\maketitle

\begin{abstract}
	Asymmetric image retrieval is a task that seeks to balance retrieval accuracy and efficiency by leveraging lightweight and large models for the query and gallery sides, respectively. The key to asymmetric image retrieval is realizing feature compatibility between different models. Despite the great progress, most existing approaches either rely on classifiers inherited from gallery models or simply impose constraints at the instance level, ignoring the structure of embedding space. In this work, we propose a simple yet effective structure similarity preserving method to achieve feature compatibility between query and gallery models. Specifically, we first train a product quantizer offline with the image features embedded by the gallery model. The centroid vectors in the quantizer serve as anchor points in the embedding space of the gallery model to characterize its structure. During the training of the query model, anchor points are shared by the query and gallery models. The relationships between image features and centroid vectors are considered as structure similarities and constrained to be consistent. Moreover, our approach makes no assumption about the existence of any labeled training data and thus can be extended to an unlimited amount of data. Comprehensive experiments on large-scale landmark retrieval demonstrate the effectiveness of our approach. Our code is released at: \url{https://github.com/MCC-WH/SSP}.
\end{abstract}


\section{Introduction}
\label{sec:intro}
\begin{figure}[t]
	\begin{center}
		\includegraphics[width=1\linewidth]{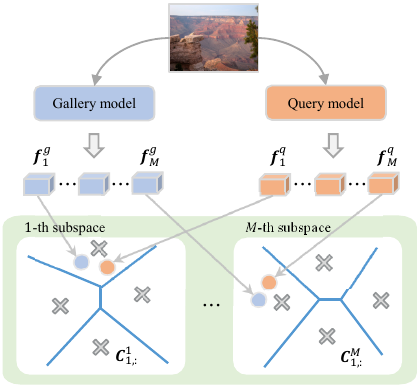}
	\end{center}
	\caption{Illustration of our structure similarity preserving method for \emph{asymmetric image retrieval}. $\bm{f}^q_{i}$ (orange) and $\bm{f}^g_{i}$ (blue) denote the $i$-th sub-vector of features from the lightweight query model and the large gallery model, respectively. $\bm{C}^{M}_{i,:}$ denotes the $i$-th centroid vector, \ie, anchor point, in the $M$-th subspace, which is generated by product quantization. During the learning of the query model, a training image is first embedded by query and gallery models, respectively. Then, we constrain the consistency of the structure similarities between these features and the anchor points to achieve feature compatibility between different models.}
	\label{fig:intro}
\end{figure}

In recent years, deep learning-based visual search methods~\cite{DSM,babenko2014neural,Wu_2021_ICCV,RMAC,DEKD,RPSBIR,9693256,8496794,7588055,9219219,7054551,8755330} have achieved great success. In a typical visual search system, a deployed deep representation model is used to embed both query and gallery images into a discriminative embedding space. Usually, the embedding features of the large-scale gallery set are extracted and indexed in advance. During the retrieval stage, query features are extracted online and the retrieval is performed by ranking the distance, \eg, Euclidean distance or cosine similarity, between the gallery features and the input query features.

Conventional methods for image retrieval, as described in previous literature~\cite{HOW,radenovic2018fine}, utilize a symmetric image retrieval approach, in which the same deep representation model is used to embed both query and gallery images~\cite{AML,HVS}.
To obtain high retrieval accuracy, a large powerful model is usually deployed, which is computationally expensive. 
However, in some real-world applications, gallery images undergo feature extraction offline on resource-rich servers, while query images are processed on resource-constrained end devices, \eg, mobile phones. 
Due to computational resource constraints, it is difficult to deploy the same large model on these end devices. Lightweight models are better choices due to their low time latency and resource footprint. 
Thanks to compatible feature learning~\cite{BCT,ijcai2022p225}, it is feasible to embed query and gallery images with lightweight and large models separately. 
This allows enjoying the excellent feature extraction capability of the large model on the server side while maintaining low resource consumption on the query side. 
Such an asymmetric setup is denoted as \emph{asymmetric image retrieval} in HVS~\cite{HVS} and AML~\cite{AML}.

As for \emph{asymmetric image retrieval}, it is crucial to ensure that features encoded by the lightweight query model and the large powerful gallery model are compatible with each other. To this end, a straightforward solution is to constrain the features encoded by the two models to be identical, which has shown effectiveness in AML~\cite{AML}. 
Another approach, adopted by BCT~\cite{BCT}, inherits the classifier of the gallery model to guide the query model. 
Recently, CSD~\cite{CSD} has considered both feature imitation and neighbor relationship preservation but requires performing multiple retrievals during the training of the query model. 
Besides, some other works further design advanced restrictions~\cite{LCE,Hot,zhang2022darwinian,9939072,Wu_Chen_Lou_Bai_Bai_Deng_Duan_2022} or advanced network structures~\cite{HVS}. 
However, all these methods impose constraints only at the instance level or require multiple online retrievals to acquire the nearest neighbor structure, failing to fully preserve the structure of embedding space during the query model learning.

To address above issues, we propose a novel structure similarity preserving approach for ensuring feature compatibility between query and gallery models, which is shown in Figure~\ref{fig:intro}. 
Our proposed approach involves the extraction of features from an independent dataset using the gallery model, followed by the training of a product quantizer (PQ) in an offline manner. 
The centroid vectors of the quantizer are then used as anchor points in the embedding space of the gallery model.
During query model training, each training image is embedded into features by both query and gallery models, and these features are converted into structure similarities by calculating similarity against anchor points. 
Our method then constrains the consistency of two structure similarities to optimize the query model. By sharing anchor points between query and gallery models, 
our method allows the features encoded by two models to align with each other while preserving the embedding space structure.

Compared to previous methods, our approach has two unique advantages. 
First, we transform features into structure similarities instead of directly performing feature regression. 
This enables the query model to ignore unimportant feature ``details" that are challenging to regress, thereby avoiding over-fitting. 
Additionally, the centroid vectors of the product quantizer encode the structure information of the gallery embedding space. 
By sharing these centroids, the embedding spaces of the query and gallery models are closely aligned so that their features are mutually interpretable. 
Second, our approach leverages the gallery model to derive the structure similarities of the training images, which are further adopted as pseudo-labels to optimize the query model. This eliminates the need for manual annotation of the training dataset, making our approach adaptable and scalable to a variety of real-world scenarios.

To evaluate our method, experiments are conducted on the Revisited Oxford and Paris datasets, with extra 1M distractor images further added for large-scale experiments. 
Comprehensive experiments with ablation study demonstrate the effectiveness of our method, which achieves the best results compared to state-of-the-art methods.

\section{Related Work}
\subsection{Image Retrieval} 
Given a large corpus, image retrieval aims to efficiently identify images that contain the same object or content as the query image based on feature similarities. Most early retrieval systems rely on bag-of-words representations~\cite{philbin2007object,sivic2003video,zhou2010spatial,ASMK} with large vocabularies and inverted indexes. In addition, methods for aggregating local features~\cite{lowe2004distinctive,bay2006surf} have been explored, including Fisher vectors~\cite{fisher_vector}, VLAD~\cite{jegou2011aggregating}, and ASMK~\cite{ASMK} that produce global descriptors capable of scaling to large databases. To further improve retrieval accuracy, various re-ranking techniques such as spatial verification~\cite{philbin2007object,zhou2010spatial}, query expansion~\cite{QE}, and diffusion~\cite{fangyuanqiang} are further adopted as post-processing steps. Recently, deep learning-based approaches have emerged as promising solutions by framing image retrieval as a metric learning task. Several loss functions~\cite{AP,ArcFace}, pooling methods~\cite{Babenko_2015_ICCV,CROW,radenovic2018fine,RMAC}, and training datasets~\cite{weyand2020google,radenovic2018revisiting} are proposed to enhance the deep representation model. 

Although a lot of efforts have been made, optimal retrieval systems typically deploy a large powerful model to process both queries and galleries, which is unaffordable on some resource-constrained end devices. In this work, we focus on \emph{asymmetric image retrieval}, where the query (user) side deploys a lightweight model, while the gallery side deploys a large powerful one.

\begin{figure*}[t]
	\begin{center}
		\includegraphics[width=1.0\linewidth]{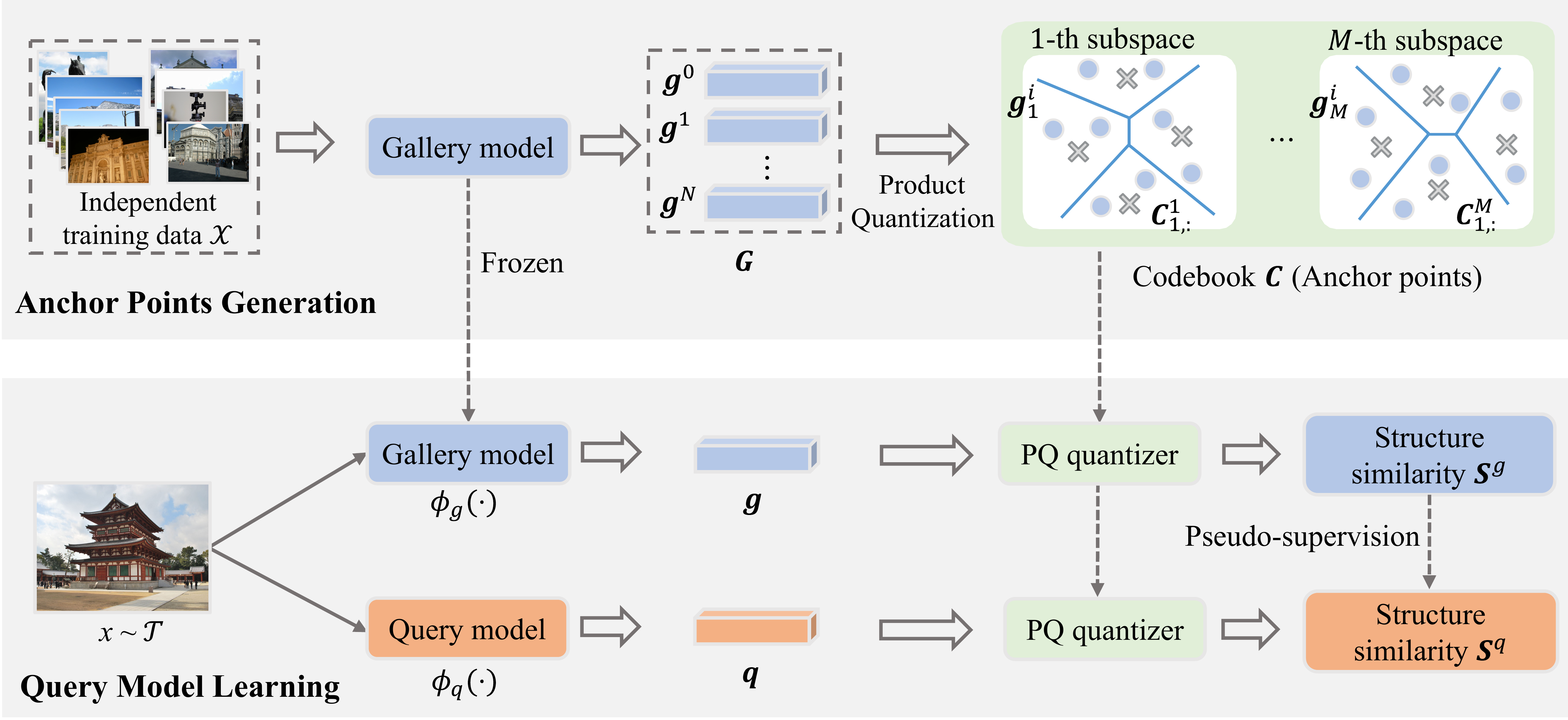}
	\end{center}
	\caption{An overview of our framework. A well trained gallery model $\phi_g(\cdot)$ is first applied to extract the image features $\left[\bm{g}^1,\bm{g}^2,\cdots,\bm{g}^N\right]$ of the training data $\gX$. Then, these features are utilized to train a product quantizer~\cite{PQ} (Section~\ref{sec:anchor_points}), whose codebook $\bm{C}$ serves as the anchor points in the embedding space of $\phi_g(\cdot)$. During query model learning, an image $x$ of training dataset $\gT$ is mapped into two feature vectors $\bm{g}$ and $\bm{q}$ by the query and gallery models, respectively. Then, the similarities between the query/gallery feature ($\bm{q}/\bm{g}$) and the anchor points are regarded as structure similarities ($\bm{S}^q/\bm{S}^g$). Finally, we constrain the consistency between structure similarities $\bm{S}^q$ and $\bm{S}^g$ to optimize the query model (Section~\ref{sec:structure_similarity_consistence}). This ensures the embedding spaces of query and gallery models are well aligned, which is essential for \emph{asymmetric retrieval}.
	}
	\label{fig:framework}
\end{figure*}

\subsection{Feature Compatible Learning} 
It is essential for \emph{asymmetric image retrieval} to encode new (query) features to
be interoperable with old (gallery) features. BCT~\cite{BCT} first introduces the problem of \emph{backward-compatible learning} and proposes to inherit the classifier of the gallery model for query model learning. AML~\cite{AML} performs asymmetric metric learning with the embeddings of anchor and positive/negative samples extracted by query and gallery models, respectively. In HVS~\cite{HVS}, both model parameters and architectures are considered simultaneously, and a compatibility-aware neural architecture search method is proposed to search for the optimal query model architecture. LCE~\cite{LCE} proposes a new classifier boundary loss to further improve feature compatibility. However, all these methods impose constraints at the instance level without considering the second-order structural information. The most relevant method to us is CSD~\cite{CSD}. It employs a strategy that incorporates both first-order feature similarity and second-order nearest-neighbor similarity between the gallery and query models. The primary objective is to utilize the gallery model to retrieve the top $k$ nearest neighbor samples from the training dataset. Subsequently, it enforces constraints on the similarity between the feature of the query model and these nearest neighbors to maintain consistency with that of the gallery model. However, it requires multiple time-consuming retrievals during each training iteration to obtain the neighbors of each sample, and using only data samples may not fully capture the structure of the embedding space.

Differently, our method generates a large number of anchor points in the embedding space with product quantization~\cite{PQ}. These anchor points are densely distributed in the embedding space of the gallery model, which carve its structure more delicately. Then, the gallery feature is converted into structure similarity by computing similarities against anchor points, which further serve as the pseudo-label to optimize the query model. Notably, the anchor points are shared by both models, so that feature compatibility is achieved.

\subsection{Lightweight Network}
With the evolution of the model architecture, deep convolutional neural networks (CNNs) have made tremendous advances in various computer vision tasks. In real-world applications, computational complexity is another important consideration in addition to accuracy. Typically, it is expected to achieve the best accuracy with a limited computational budget, which is determined by the target computing platforms. The immediate need to deploy high-performance deep neural networks on a range of resource-constrained end devices has motivated a series of studies on efficient model design, including SqueezeNets~\cite{iandola2016squeezenet}, MobileNets~\cite{MobileNets,Mobilenetv2}, ShuffleNets~\cite{ zhang2018shufflenet,shufflenetv2} and EfficientNets~\cite{Efficientnet}. All these methods aim at designing lightweight architecture to achieve better speed-accuracy trade-offs. 

In this work, we focus on \emph{asymmetric image retrieval}, where query features are extracted on some resource-constrained end devices. Our approach employs the various lightweight models mentioned above as query models.

\subsection{Knowledge Transfer}
Knowledge transfer aims at learning a student model by transferring knowledge from a pretrained teacher model. It is first introduced by Hinton~\etal~\cite{hinton2015},
where the student model learns from real labels and soft predicted class logits by the teacher. FitNet~\cite{fitnets} distills knowledge through intermediate features and Euclidean distance is used to measure the distance between them. After that, PKT~\cite{PKT} models the knowledge of the teacher model as a probability distribution and uses KL divergence to measure the distance. In RKD~\cite{RKD} and DARK~\cite{DARK}, geometric relationships between multiple examples, such as angles and distances, are used as knowledge to guide students learning. CRD~\cite{CRD} combines contrastive learning and knowledge distillation and uses contrastive loss to transfer knowledge between different modalities. Other approaches use multi-stage information to transfer knowledge. AT~\cite{AT} uses multi-layer attention maps to transfer knowledge. FSP~\cite{FSP} generates FSP matrices from layer features and uses them to guide the learning process of a small model.

However, these methods only transfer knowledge between models but do not consider feature compatibility between them. Thus, they fail to meet the needs of \emph{asymmetric image retrieval}. In our approach, centroid vectors of a product quantizer, which is trained using the image features extracted by the gallery model, serve as the anchor points for both models. Thus, the embedding spaces of the query and gallery model are constrained to be aligned during knowledge transfer.

\section{Our Approach}
In this section, we first give a formulation of asymmetric image retrieval. After that, we elaborate our structure similarity preserving framework.

\subsection{Problem Formulation}
Let $\phi_g(\cdot)$ and $\phi_q(\cdot)$ denote the gallery and query models, respectively. For a visual retrieval system, the gallery model $\phi_g(\cdot)$ is first trained and then used to map the gallery images $\gR$ into feature vectors. During testing, the query model $\phi_q(\cdot)$ processes queries $\gQ$, and the retrieval is reduced to the nearest neighbor search in the embedding space. Some evaluation metric, \eg, mean Average Precision (mAP), is used to evaluate the performance of a retrieval system, which is abbreviated as $\gM(\phi_q(\cdot),\phi_g(\cdot))$ for simplicity. In a conventional symmetric retrieval system, the query model is usually the same as the gallery model, \ie, $\phi_q(\cdot)=\phi_g(\cdot)$. It typically deploys a large powerful model to achieve high retrieval accuracy, which cannot be satisfied in resource-constrained scenarios. 

As for an asymmetric retrieval system, the model $\phi_g(\cdot)$ is well-trained and fixed. To facilitate resource-constrained application scenarios, it needs to learn a compatible lightweight query model $\phi_q(\cdot)$ which is significantly smaller than $\phi_g(\cdot)$ in terms of parameter size and computational complexity. The core of \emph{asymmetric image retrieval} is that the feature embeddings of query and gallery models are mutually interpretable. In other words, we expect that \emph{asymmetric image retrieval} achieves a retrieval accuracy similar to that of \emph{symmetric retrieval}, \ie, $\mathcal{M}(\phi_q(\cdot), \phi_g(\cdot)) \approx \mathcal{M}(\phi_g( \cdot),\phi_g(\cdot))$ so that the balance between performance and efficiency is achieved.

\subsection{Structure Similarity Preservation Learning}\label{sec:overview}
In this work, we propose a structure similarity preserving framework, which is shown in Figure~\ref{fig:framework}. A product quantizer is first trained with the features extracted by the gallery model. The centroids of the quantizer serve as the anchor points to characterize the space structure. During the training of the query model, the gallery model is frozen. Each training sample is mapped into two embeddings by the query and gallery models, respectively. Then, two embeddings are converted to structure similarities by calculating the similarities against the centroids. Finally, our method restricts the consistency between two structure similarities to optimize the query model. Since the anchor points generated by the gallery model are shared with the query model, their embedding space is well-aligned after training.

\subsubsection{Anchor Points Generation}\label{sec:anchor_points}
To achieve a comprehensive characterization of the embedding space, our approach requires selecting representative anchor points in the embedding space of the gallery model. These anchor points are fixed reference points in the embedding space which are used to convert query and gallery features into structure similarities. A straightforward approach is to use flat $k$-means clustering to generate a series of anchor points. However, our method requires a large number of anchor points to delicately characterize the space structure. If $k$-means clustering is adopted, the required training samples and computational complexity are several times the number of centroids. When the number of centroids is large, the cost of clustering is unaffordable. To this end, our approach employs product quantization (PQ)~\cite{PQ} to efficiently expand the number of anchor points at a lower cost. 

Suppose there exists some training data $\mathcal{X} = \left\{x_1,x_2,\cdots,x_N\right\}$ for anchor points generation. Since the gallery model $\phi_g(\cdot)$ is frozen, we first employ it to extract the features $\bm{G} = \left[\bm{g}^1;\bm{g}^2;\cdots;\bm{g}^N\right] \in \mathbb{R}^{N\times d}$ of images in $\mathcal{X}$ offline:
\begin{eqnarray}
	\label{equ:extract_offline}
	\bm{g}^i=\phi_g(x_i) \in \mathbb{R}^d,~i=1,2,\cdots,N.
\end{eqnarray} Then, each feature vector $\bm{g}^i \in \bm{G}$ is split into $M$ distinct sub-vectors $u_j(\bm{g}^i)\in \mathbb{R}^{d^{*}},~j=1,2,\cdots,M$:
\begin{eqnarray}\label{equ:M_split}
	\begin{aligned}
		\underbrace{\bm{g}_1^i,\cdots,\bm{g}_{d^{*}}^i}_{u_1(\bm{g}^i)},\cdots,\underbrace{\bm{g}_{d-d^{*}+1}^i,\cdots,\bm{g}_{d}^i}_{u_M(\bm{g}^i)},
	\end{aligned}
\end{eqnarray}
where $\bm{g}^{i}_j$ denotes the $j$-th feature dimension of $\bm{g}^i$, $d^{*} = d/M$ and $d$ is a multiple of $M$. After that, we perform $k$-means clustering on each sub-vector set $\left[u_j(\bm{g}^1);u_j(\bm{g}^2);\cdots;u_j(\bm{g}^N)\right] \in \mathbb{R}^{N \times d^{*}},~j=1,2,\cdots,M,$ individually to obtain the corresponding sub-codebook $\bm{C}^j \in \mathbb{R}^{K\times d^{*}}$, where $K$ is the number of centroids. Then, the anchor points in the gallery space are defined as the Cartesian product of sub-codebooks:
\begin{eqnarray}\label{equ:Cartesian_product}
	\begin{aligned}
		\bm{C} = \bm{C}^1 \times \bm{C}^2 \times \cdots \times \bm{C}^M\in \mathbb{R}^{K^M\times d},
	\end{aligned}
\end{eqnarray}
in which any centroid vector is formed by concatenating $M$ different sub-centroid vectors. 

Compared with $k$-means clustering, PQ has two distinctive advantages. First, it is easy to generate a large number of anchor points $\bm{C}$. The total number of anchor points is $K^M$. Second, instead of storing the huge anchor points directly, it only needs to store $M \times K$ sub-centroids. During training, we also adopt the splitting mechanism to calculate the similarity by segments, instead of directly computing the similarities between feature vectors and all anchor points, which greatly reduces our training overhead. The complete learning procedure is summarized in Algorithm~\ref{alg:anchor}.

\begin{algorithm}[ht]
	\DontPrintSemicolon
	\KwInput{Training data $\mathcal{X} = \left\{x_1,x_2,\cdots,x_N\right\}$; gallery model $\phi_g(\cdot)$; number of subvectors $M$; number of centroids per subvector $K$}
	\KwOutput{Anchor points $\bm{C} = \bm{C}^1 \times \bm{C}^2 \times \cdots \times \bm{C}^M\in \mathbb{R}^{K^M\times d}$}
	\For { each image $x_i$ in traning data $\mathcal{X}$}
	{
		Extract image feature with gallery model according to Equation~(\ref{equ:extract_offline});\\
		Split image feature into $M$ sub-vectors $u_j(\bm{g}^i)\in \mathbb{R}^{d^{*}},~j=1,2,\cdots,M$ according to Equation~(\ref{equ:M_split});\\
	}
	
	\For { each sub-vectors set $\left[u_j(\bm{g}^1);u_j(\bm{g}^2);\cdots;u_j(\bm{g}^N)\right] \in \mathbb{R}^{N \times d^{*}}$}
	{
		Perform $k$-means clustering with $K$ centroids; \\
		Obtain corresponding sub-codebook $\bm{C}^j \in \mathbb{R}^{K\times d^{*}}$;\\
	}
	\caption{Anchor points generation}
	\label{alg:anchor}
\end{algorithm}

\subsubsection{Query Model Learning}\label{sec:query_model_learning}
\noindent \textbf{Structure Similarity Calculation}.\label{sec:structure_similarity_caculate}
During the query model learning, the feature vectors of query and gallery models are first converted into structure similarities by calculating the similarities against anchor points. Given an image $x$ in training dataset $\mathcal{T}$. Let $\bm{g}$ and $\bm{q}$ be its feature vectors extracted by the gallery and query models, respectively:
\begin{eqnarray}
	\label{equ:training_extract}
	\bm{g}=\phi_g(x) \in \mathbb{R}^d,~\bm{q}=\phi_q(x) \in \mathbb{R}^d.
\end{eqnarray}

We first split them into $M$ sub-vectors:
\begin{eqnarray}\label{equ:M_segment}
	\begin{aligned}
		\bm{g} & \rightarrow u_1(\bm{g}), u_2(\bm{g}),\cdots,u_M(\bm{g}), \\
		\bm{q} & \rightarrow u_1(\bm{q}), u_2(\bm{q}),\cdots,u_M(\bm{q}). \\
	\end{aligned}
\end{eqnarray}
Then, we calculate the structure similarities $\bm{S}^{g}_{i,:}$ and $\bm{S}^{q}_{i,:}$ by computing for each sub-vectors $u_i(\bm{g})$ and $u_i(\bm{q})$ the similarities against the corresponding $K$ centroid vectors in the pretrained quantizer:
\begin{eqnarray}\label{equ:S}
	\begin{aligned}
		\bm{S}^{g}_{i,:} = \left[s(u_i(\bm{g}),\bm{C}_{1,:}^i),\cdots,s(u_i(\bm{g}),\bm{C}_{K,:}^i)\right] \in \mathbb{R}^{K}, \\
		\bm{S}^{q}_{i,:} = \left[s(u_i(\bm{q}),\bm{C}_{1,:}^i),\cdots,s(u_i(\bm{q}),\bm{C}_{K,:}^i)\right] \in \mathbb{R}^{K},
	\end{aligned}
\end{eqnarray}
where $\bm{C}_{i,:}^j$ denotes the $i$-th centroid vector in the $j$-th subspace and $s(\cdot,\cdot)$ is the similarity metric. In this work, cosine similarity is considered, and $s(\cdot,\cdot)$ is formulated as:
\begin{eqnarray}\label{equ:cosine}
	\begin{aligned}
		s(u_i(\bm{g}),\bm{C}_{i,:}^j) = \frac{\bm{C}_{i,:}^ju_i(\bm{g})^T}{\| \bm{C}_{i,:}^j\|_2\|u_i(\bm{g})\|_2}.
	\end{aligned}
\end{eqnarray}

Finally, we impose consistency constraints $\mathcal{L}_c$ on the structure similarities $\bm{S}^{g}$ and $\bm{S}^{q}$ to optimize $\phi_q(\cdot)$ so that the feature embedding $\bm{q}$ shares the same structure similarity as $\bm{g}$ in the embedding space of the gallery model. Notably, the anchor points are shared between the query and gallery models, thus their embedding spaces are well aligned.

\noindent \textbf{Structure Similarity Preserving Constraint}.\label{sec:structure_similarity_consistence}
For \emph{asymmetric image retrieval}, a desirable query model $\phi_q(\cdot)$ not only maintains feature compatibility but also preserves the structure similarity of $\bm{g}$ in the embedding space of gallery model. To this end, our method constrains the consistency between two structure similarity $\bm{S}^{g}_i$ and $\bm{S}^{q}_i$ for the corresponding sub-vector pair $u_i(\bm{g})$ and $u_i(\bm{q})$. 

Specifically, Kullback–Leibler (KL) divergence is adopted to measure the distance between $\bm{S}^{g}_{i,:}$ and $\bm{S}^{q}_{i,:}$. First, $\bm{S}_i^{\bm{g}}$ is converted into the form of probability distribution:
\begin{eqnarray}\label{equ:probalility}
	\begin{aligned}
	\bm{p}_{i,:}^g \! \!= \!\!\left[\frac{\exp\left(\bm{S}_{i,1}^g/\tau_g\right)}{\sum_{l=1}^{K}\exp\left(\bm{S}_{i,l}^g/\tau_g\right)},\cdots,\frac{\exp\left(\bm{S}_{i,K}^g/\tau_g\right)}{\sum_{l=1}^{K}\exp\left(\bm{S}_{i,l}^g/\tau_g\right)}\right],\!
	\end{aligned}
\end{eqnarray}
where $\tau_g$ is a temperature value used for controlling the sharpness of the assignments.
Similarly, the probability distribution corresponding to the $i$-th subvector of the query feature $\bm{q}$ is formulated as: 
\begin{eqnarray}\label{equ:probalility_stu}
	\begin{aligned}
		\bm{p}_{i,:}^q \!\!=\!\!\left[\frac{\exp\left(\bm{S}_{i,1}^q/\tau_q\right)}{\sum_{l=1}^{K}\exp\left(\bm{S}_{i,l}^q/\tau_q\right)},\cdots,\frac{\exp\left(\bm{S}_{i,K}^q/\tau_q\right)}{\sum_{l=1}^{K}\exp\left(\bm{S}_{i,l}^q/\tau_q\right)}\right].\!
	\end{aligned}
\end{eqnarray}

Then, the structure similarity preserving constraint is defined as the KL divergence between two probabilities over the same sub-centroid vectors:
\begin{eqnarray}
	\label{equ:KL}
	\mathcal{L}_{\text{KL}}^i = \text{KL}(\bm{p}_{i,:}^{g} || \bm{p}_{i,:}^{q}) = \sum_{l=1}^{K}\bm{p}_{i,l}^{g}\log{\frac{\bm{p}_{i,l}^{g}}{\bm{p}_{i,l}^{q}}}, 
\end{eqnarray}
which consists of the cross-entropy of $\bm{p}_i^{g}$ and $\bm{p}_i^{q}$, and the entropy of $\bm{p}_i^{g}$. The latter is independent of the feature of the query model and thus does not affect the training. The final objective function is defined as the summation of all consistency losses corresponding to the $M$ distinct sub-vectors:
\begin{eqnarray}
	\label{equ:final}
	\mathcal{L}_{\text{final}} = \sum_{i=1}^{M}\mathcal{L}_{\text{KL}}^i,
\end{eqnarray}
which is used for optimizing query model end to end. 

\begin{algorithm}[ht]
	\DontPrintSemicolon
	\KwInput{Training set $\mathcal{T}$; well-trained gallery model $\phi_g(\cdot)$; random initialized query model $\phi_q(\cdot)$; anchor points $\bm{C}$}
	\KwOutput{Query model $\phi_q(\cdot)$ compatible with $\phi_g(\cdot)$}
	\For { each image $x$ in traning set $\mathcal{T}$}
	{
		Extract image feature with gallery and query models according to Equation~(\ref{equ:training_extract});\\
		Split $\bm{g}$ and $\bm{q}$ into $M$ sub-vectors according to Equation~(\ref{equ:M_segment});\\
		Calculate the structure similarities $\bm{S}^{g}_{i,:}$ and $\bm{S}^{q}_{i,:}$ according to Equation~(\ref{equ:S});\\
		Impose consistency constraints $\mathcal{L}_c$ on the structure similarities $\bm{S}^{g}$ and $\bm{S}^{q}$ to optimize $\phi_q(\cdot)$ according to Equation~(\ref{equ:L_c});\\
	}
	\caption{Query model learning}
	\label{alg:query_model}
\end{algorithm}

\noindent \textbf{Soft-assignment \emph{vs}\onedot Hard-assignment}.
The centroids of the product quantizer serve as the anchor points in the embedding space of the gallery model. By quantizing the feature, we convert the feature regression into an assignment prediction task.
When setting temperature $\tau_g =0 $, the probability $\bm{p}_{i,:}^{g}$ in Equation~(\ref{equ:probalility}) will be a one-hot vector with the only 1 at index $j = \arg\max_{j}(\bm{S}^g_{i,j})$. Thus, Equation~(\ref{equ:probalility}) is simplified as
\begin{eqnarray}
	\label{equ:L_c}
	\mathcal{L}_{\text{KL}}^i = \sum_{l=1}^{K}\bm{p}_{i,l}^{g}\log{\frac{\bm{p}_{i,l}^{g}}{\bm{p}_{i,l}^{q}}} = \log{\frac{1}{\bm{p}_{i,j}^{q}}}.
\end{eqnarray}

Optimizing this loss encourages the query model to regress the anchor point, to which feature $\bm{g}$ is quantized. It avoids the query model regressing the feature ``details'' of the gallery model. However, the relationships between the feature vector and anchor points carry discriminative knowledge, and simply ignoring them may lead to inferior performance. Thus, we set $\tau_g > 0$ to use the soft assignments as the prediction target. The overall learning process of the query model is summarized in Algorithm~\ref{alg:query_model}.

\begin{table}[t]
	\begin{center}
		\small
		\setlength{\extrarowheight}{1.0pt}
		\setlength{\tabcolsep}{7.0pt}
		\begin{tabular}{ccccc} 
			\toprule
			\mr{2}{\tabincell{c}{\Th{Query}\\\Th{Model} $\phi_q(\cdot)$}} & \mc{2}{\Th{Flops (G)}} & \mc{2}{\Th{Param(M)}} \\ \cmidrule(l){2-3} \cmidrule(l){4-5}
			& ABS & $\%$ & ABS & $\%$ \\ 
			\midrule
			ResNet101~\cite{resnet} & 42.85  & 100.0  & 42.50 & 100.0 \\
			\midrule
			ShuffleNetV2 ($0.5\times$)~\cite{shufflenetv2} & 0.84  & 1.96  & 2.44 & 5.74 \\
			ShuffleNetV2~\cite{shufflenetv2} & 1.44  & 3.36  & 3.35 & 7.88 \\
			MobileNetV2~\cite{Mobilenetv2} & 2.50  & 5.83  & 4.85 & 11.41 \\
			EfficientNetB0~\cite{Efficientnet} & 2.86  & 6.67  & 6.63 & 15.60 \\
			EfficientNetB1~\cite{Efficientnet} & 3.92  & 9.15  & 9.13 & 21.49 \\
			EfficientNetB2~\cite{Efficientnet} & 4.50  & 10.51  & 10.58 & 24.90 \\
			EfficientNetB3~\cite{Efficientnet} & 6.24  & 14.57  & 13.84 & 32.56 \\
			\bottomrule
		\end{tabular}
	\end{center}\vspace{-1.0em}
	\caption{The computational complexity and the number of parameters of all lightweight query models adopted in this work are listed. ABS: absolute number. $\%$: relative to ResNet101~\cite{resnet}. ($0.5\times$) denotes a model with $0.5\times$ width. \Th{Flops} are calculated when the input image size is $362 \times 362$.\vspace{-1.0em}}
	\label{tab:net_params}
\end{table}

\section{Experiments}
\subsection{Experimental Setup}
\noindent \tb{Dataset}. SfM-120k~\cite{radenovic2018fine} and the clean version of Google landmark v2 (GLDv2)~\cite{weyand2020google} are adopted as training set $\gT$. SfM-120k includes images selected
from 3D reconstructions of landmarks and city scenes. Following the common setting~\cite{AML}, we use 91,642 images from 551 3D models for training and the remaining images of 162 3D models for validation. The clean version of GLDv2~\cite{weyand2020google} consists of 1,580,470 images from 81,313 categories. We randomly select $80\%$ images as the training set and let the rest as the validation set. 
$\gR$1M~\cite{radenovic2018revisiting} is used as extra images $\mathcal{X}$ for anchor points generation. It is collected from Yahoo Flickr Creative Commons 100 Million (YFCC100m) dataset~\cite{thomee2016yfcc100m} and contains 1M distractor images.

\noindent\tb{Query and Gallery Models}. ResNet101 trained by DELG~\cite{DELG} and GeM~\cite{radenovic2018fine} are deployed as gallery models, which are denoted as R101-DELG and R101-GeM in this work, respectively. As for lightweight query models, ShuffleNets~\cite{shufflenetv2}, MobileNets~\cite{Mobilenetv2} and EfficientNets~\cite{Efficientnet} are chosen. To adapt the model for image retrieval tasks, only the feature extractor of the model is kept and the other layers are both removed. Then, GeM pooling~\cite{radenovic2018fine} is applied on the last convolutional feature map, followed by another whitening layer, which is implemented by a fully-connected layer. The whitening layer is initialized in the embedding space of the gallery models and kept frozen during the training of the query model. In Table~\ref{tab:net_params}, we list the number of parameters and the computational complexity (in \Th{Flops}) of the lightweight models adopted in this work.

\noindent \tb{Evaluation Datasets and Metrics}. The revisited Oxford5k~\cite{philbin2007object} and Paris6k~\cite{philbin2008lost} datasets are used for evaluation, which are denoted as $\gR$Oxf and $\gR$Par~\cite{radenovic2018revisiting}. All datasets describe
specific landmarks of buildings under a variety of different observation conditions, each with 70 query images, and 4,993 and 6,322 gallery images, respectively. We follow the common setting~\cite{AML} to report mAP under the Medium and Hard settings for two datasets. Large-scale experiment results are further reported with the $\gR$1M (1M distractor images)~\cite{radenovic2018revisiting} dataset added to the database.

\begin{table*}[tb]
	\begin{center}
		\normalsize
		\setlength{\extrarowheight}{1.0pt}
		\setlength{\tabcolsep}{2pt}
		\resizebox{\linewidth}{!}{
		\begin{tabular}{lcccccccccccc}
			\toprule
			\multicolumn{2}{c}{ \multirow{2}*{\Th{Method}}}& \multirow{2}*{\tabincell{c}{ \Th{Query}\\\Th{Model}}}& \multirow{2}*{ \tabincell{c}{ \Th{Gallery}\\\Th{Model}}} & \mr{2}{\tabincell{c}{\Th{Training}\\\Th{Set} $\mathcal{T}$}} & \multicolumn{4}{c}{\Th{Medium}} &\multicolumn{4}{c}{\Th{Hard}}\\
			\cmidrule(l){6-9}  \cmidrule(l){10-13}
			\multicolumn{5}{c}{} & $\mathcal{R}$Oxf &
			$\mathcal{R}$Oxf+$\mathcal{R}$1M &
			$\mathcal{R}$Par &
			$\mathcal{R}$Par+$\mathcal{R}$1M &
			$\mathcal{R}$Oxf &
			$\mathcal{R}$Oxf+$\mathcal{R}$1M &
			$\mathcal{R}$Par &
			$\mathcal{R}$Par+$\mathcal{R}$1M  \\
   
			\midrule

			\multicolumn{2}{l}{\color{gray}{GeM$\dagger$}~\cite{radenovic2018fine}}&\color{gray}{R101-GeM}&\color{gray}{R101-GeM}& \mr{3}{\color{gray}{SfM-120k}}& \tb{\color{gray}{65.43}}  & \tb{\color{gray}{45.23}} & \tb{\color{gray}{76.75}} & \tb{\color{gray}{52.34}} & \tb{\color{gray}{40.13}} & \tb{\color{gray}{19.92}} & \tb{\color{gray}{55.24}} & \tb{\color{gray}{24.77}} \\

            \multicolumn{2}{l}{\color{gray}{GeM$\dagger$}~\cite{radenovic2018fine}}&\color{gray}{EfficientNetB3}&\color{gray}{EfficientNetB3}& & \color{gray}{54.22}  & \color{gray}{37.10} & \color{gray}{71.21} & \color{gray}{44.67} & \color{gray}{27.53} & \color{gray}{17.49} & \color{gray}{48.00} & \color{gray}{18.45} \\
            
            \multicolumn{2}{l}{\color{gray}{GeM$\dagger$}~\cite{radenovic2018fine}}&\color{gray}{MobileNetV2}&\color{gray}{MobileNetV2}& & \color{gray}{58.81} & \color{gray}{40.02} & \color{gray}{67.87} & \color{gray}{42.25} & \color{gray}{33.41} & \color{gray}{17.71} & \color{gray}{40.97} & \color{gray}{16.59} \\
            
			\cdashlinelr{1-13}	
    		\multicolumn{2}{l}{\color{gray}{DELG$\dagger$}~\cite{DELG}}&\color{gray}{R101-DELG}&\color{gray}{R101-DELG} & \mr{3}{\color{gray}{GLDv2}} &\tb{\color{gray}{78.55}}  &\tb{\color{gray}{66.02}} &\tb{\color{gray}{88.58}} &\tb{\color{gray}{73.65}} &\tb{\color{gray}{60.89}} &\tb{\color{gray}{41.75}} &\tb{\color{gray}{76.05}} &\tb{\color{gray}{51.46}} \\

                \multicolumn{2}{l}{\color{gray}{DELG$\dagger$}~\cite{DELG}}&\color{gray}{EfficientNetB3}&\color{gray}{EfficientNetB3} &  & \color{gray}{66.64} & \color{gray}{49.67} & \color{gray}{81.78} & \color{gray}{61.10} & \color{gray}{43.82} & \color{gray}{24.89} & \color{gray}{63.90} & \color{gray}{32.34} \\
                
                \multicolumn{2}{l}{\color{gray}{DELG$\dagger$}~\cite{DELG}}&\color{gray}{MobileNetV2}&\color{gray}{MobileNetV2} &  & \color{gray}{62.42} & \color{gray}{42.21} & \color{gray}{77.91} & \color{gray}{55.09} & \color{gray}{36.56} & \color{gray}{18.64} & \color{gray}{57.96} & \color{gray}{28.81} \\

			\midrule
			\multicolumn{2}{l}{~Contr$^{*}$~\cite{AML}}&\multirow{4}*{MobileNetV2} & \multirow{4}*{R101-GeM}&\mr{4}{SfM-120k}&47.10&18.00&61.50&28.80& 21.80&6.30&37.70&8.80\\
			
			\multicolumn{2}{l}{~Reg~\cite{AML}}&&&&49.20&26.50&65.00&34.60&23.30&7.80&40.70&12.70\\    
			
			\multicolumn{2}{l}{~CSD~\cite{CSD}}&&&& 63.59& 40.29 &76.05& 43.08 &\tb{38.51} & 17.93 & 52.67 & 17.43 \\  
			
			\multicolumn{2}{l}{~\textbf{Ours}}&&&& \tb{63.98}& \tb{41.07} &\tb{76.54}& \tb{45.40} &37.91 & \tb{19.22} & \tb{53.59} & \tb{19.02} \\  
			
			\cdashlinelr{1-13}	
			
			\multicolumn{2}{l}{~Contr$^{*}$~\cite{AML}}& \multirow{4}*{EfficientNetB3} & \multirow{4}*{R101-GeM} &\mr{4}{SfM-120k} &45.20&24.70&63.70&32.80&19.60&12.20&40.90&12.50\\
			
			\multicolumn{2}{l}{~Reg~\cite{AML}}&&&&52.90&29.70&65.20&39.00&27.80&10.40&42.40&16.00\\
			
			\multicolumn{2}{l}{~CSD~\cite{CSD}}&&&& 64.49& 43.39 &76.11 & 45.58 &39.06 & 19.12 & 53.64& 19.78 \\   
			
			\multicolumn{2}{l}{~\textbf{Ours}}&&&&  \tb{65.14}& \tb{43.95} &\tb{76.87}& \tb{48.22} &\tb{39.38} & \tb{20.01} & \tb{54.50} & \tb{20.64} \\
			\midrule
			
			\multicolumn{2}{l}{~Contr$^{*}$~\cite{AML}}& \multirow{6}*{MobileNetV2} & \multirow{6}*{R101-DELG} &\mr{6}{GLDv2}&66.42&45.76&83.13&53.10&45.99&23.34&66.79&30.24\\ 
			
			\multicolumn{2}{l}{~Reg~\cite{AML}}& && &72.75&56.03&85.81&65.23&53.07&32.21&69.96&39.29\\
			
			\multicolumn{2}{l}{~HVS\cite{BCT}}&&&&74.39&58.24&86.86&67.44&54.68&34.77&72.42&43.39\\ 
			
			\multicolumn{2}{l}{~LCE\cite{LCE}}&&&&75.45&58.03&87.24&67.30&54.95&33.88&73.03&43.01\\
			
			\multicolumn{2}{l}{~CSD~\cite{CSD}}&&&&75.94& 59.45 &87.27& 68.52  &57.51 & 36.41 & 73.45 & 44.31\\
			
			\multicolumn{2}{l}{~\textbf{Ours}}&&&& \tb{77.88}& \tb{60.26} &\tb{88.34}& \tb{70.23}  &\tb{60.05} & \tb{37.29} & \tb{75.08} & \tb{46.16} \\
   
			\cdashlinelr{1-13}	  			 		     
			
			\multicolumn{2}{l}{~Contr$^{*}$~\cite{AML}}& \multirow{6}*{EfficientNetB3} &  \multirow{6}*{R101-DELG} &\mr{6}{GLDv2}&69.45&49.70&83.81&59.36&46.19&26.49&68.15&35.24\\
			
			\multicolumn{2}{l}{~Reg~\cite{AML}}& & &&74.60&59.88&86.09&67.69&53.41&33.31&72.21&42.63\\ 
			
			\multicolumn{2}{l}{~HVS~\cite{BCT}}&&&&76.41&62.72&87.07&71.54&56.13&36.86&74.53&49.09\\
			
			\multicolumn{2}{l}{~LCE~\cite{LCE}}&&&&75.89&61.90&86.63&70.98&55.21&36.53&73.62&48.94\\
			
			\multicolumn{2}{l}{~CSD~\cite{CSD}}&&&& 77.64 & \bf{64.29} & 87.95& 72.90 & 59.32 & \tb{39.84} & 75.11 & 49.13 \\  
			
			\multicolumn{2}{l}{~\textbf{Ours}}&&&& \tb{79.46}& 63.22 & \tb{89.14}& \tb{73.07}  &\tb{62.17} & 39.05 & \tb{76.88} & \tb{49.54} \\ 
			\bottomrule
		\end{tabular}
	}
	\end{center}\vspace{-1.0em}
	\caption{mAP comparison against existing methods on the full benchmark. Black bold: best results under the same setting. $\dagger$: our re-implementation. The first six rows illustrate the performance of our method when using both large models (R101-GeM and R101-DELG) and small models (MobileNetV2 and EfficientNetB3) under a symmetrical setting.\vspace{-1.0em}}
	\label{tab:state-of-the-art}
\end{table*}

\noindent\textbf{Implementation Details}. 
Under the \emph{asymmetric image retrieval} setting, gallery models typically use very deep models (\eg, ResNet101~\cite{resnet}). It is expensive to extract the embeddings of training images during training in terms of computation and memory, especially with such large gallery models. In addition, the gallery model is not optimized during the query model learning. Therefore, our method first extracts all the embedding of training images with the large gallery models offline and caches them in memory.

When SfM-120k is adopted as the training set $\gT$, we follow the settings of AML~\cite{AML}. The query models are trained on an NVIDIA RTX 3090 GPU for 10 epochs with a batch size of 64. When GLDv2 is adopted, the image size is set to $512\times512$. Following the setting of DELG~\cite{DELG}, random cropping, random color jittering, and random horizontal flipping are used as data augmentation. We train the query models on four NVIDIA RTX 3090 GPUs with a batch size of 256 for 5 epochs. All models are optimized using Adam with an initial learning rate of $10^{-3}$ and a weight decay of $10^{-6}$. A linear decay scheduler is employed to gradually decay the learning rate to 0 when the desired number of steps is reached. $\tau_g$ and $\tau_q$ were set to 0.1 and 1.0, respectively. For the anchor points generation, the number of centroids $K$ in each subspace is set to $256$. The number of subspaces $M$ is set to 64 and 32 when R101-GeM and R101-DELG are adopted as the gallery model, respectively. 

During the testing phase, images are resized to a maximum size of $1024 \times 1024$ pixels while maintaining their original aspect ratio. Image features are extracted at three scales, namely, ${1/\sqrt{2},1,\sqrt{2}}$. We apply $L_2$ normalization to each scale independently, followed by averaging the features of the three scales and applying another $L_2$ normalization. Under the \emph{asymmetric image retrieval} setting, we leverage the lightweight query model $\phi_q(\cdot)$ to extract the features of queries and perform retrieval in the gallery, whose features are extracted by a large gallery model $\phi_g(\cdot)$.

\begin{table}[t]
	\begin{center}
		\small
		\setlength{\extrarowheight}{0.0pt}
		\setlength{\tabcolsep}{2.0pt}
            \resizebox{\linewidth}{!}{
    		\begin{tabular}{clccccc} 
    			\toprule
    			\mr{2}{\tabincell{c}{ \Th{Gallery}\\\Th{Model}}}&\mr{2}{\tabincell{c}{ \Th{Quantizer}\\\Th{Type}}} & \mr{2}{\tabincell{c}{\Th{Training}\\\Th{Set} $\mathcal{T}$}} &\mc{2}{\Th{Medium}} & \mc{2}{\Th{Hard}} \\ 
                    \cmidrule(l){4-5} \cmidrule(l){6-7}
    			& & &$\mathcal{R}$Oxf & $\mathcal{R}$Par & $\mathcal{R}$Oxf & $\mathcal{R}$Par \\ 
    			\midrule
    			\multirow{8}*{R101-GeM~\cite{radenovic2018fine}}& $k$-means$_{1,024}$ & \mr{8}{SfM-120k} & 31.1  & 51.9  & 11.6 & 23.8 \\
                    & Spectral$_{1,024}$ & & 30.1  & 54.5  & 10.3 & 25.9 \\
    			& $k$-means$_{4,096}$  & & 40.9  & 58.3  & 18.3 & 30.7 \\
                    & Spectral$_{4,096}$ & & 43.7  & 57.9  & 20.1 & 29.4 \\
    			& $k$-means$_{16,384}$ & & 46.7  & 56.9  & 21.6 & 28.5 \\
                    & Spectral$_{16,384}$& & 47.2  & 59.1  & 21.8 & 32.7 \\
    			& $k$-means$_{65,536}$ & & 52.8  & 60.8  & 26.9 & 33.2 \\
    			& PQ$_{64\|256}$ & & \tb{63.9} & \tb{76.5} & \tb{37.9} & \tb{53.5} \\
    			\midrule
    			\multirow{8}*{R101-DELG~\cite{DELG}}& $k$-means$_{1,024}$ & \mr{8}{GLDv2} & 38.3  & 56.5  & 18.9 & 36.4 \\
                    & Spectral$_{1,024}$ & & 35.8  & 54.2  & 17.1 & 37.6 \\
    			& $k$-means$_{4,096}$  & & 59.3  & 73.9  & 37.2 & 53.6 \\
                    & Spectral$_{4,096}$ & & 57.1  & 73.0  & 36.1 & 52.9 \\
    			& $k$-means$_{16,384}$ & & 63.4  & 79.0  & 39.7 & 59.7 \\
                    & Spectral$_{16,384}$& & 62.7  & 79.6  & 38.4 & 60.5 \\
    			& $k$-means$_{65,536}$ & & 67.1  & 81.1  & 46.0 & 62.4 \\
    			& PQ$_{32\|256}$ & & \tb{77.8} & \tb{88.3} & \tb{60.0} & \tb{75.0} \\
    			\bottomrule
    		\end{tabular}
            }
	\end{center}\vspace{-0.5em}
	\caption{mAP (asymmetric) comparison of \textbf{different quantizers}. MobileNetV2~\cite{Mobilenetv2} is used as query model. ``$k$-means$_{i}$'' means a flatten $k$-means quantizer with $i$ centroids. ``Spectral$_{i}$'' means that we cluster the data with $i$ centroids using spectral clustering. PQ$_{32\|256}$ denotes that we split the feature vector into 32 subvectors, with each subvector quantized to $256$ centroids. R101-GeM and R101-DELG denote the ResNet101 trained by GeM~\cite{radenovic2018fine} and DELG~\cite{DELG}, respectively.}
	\label{tab:Flatten_Kmeans}
\end{table}

\begin{table}[tb]
	\begin{center}
		\small
		\setlength{\extrarowheight}{0.0pt}
		\setlength{\tabcolsep}{2.0pt}
            \resizebox{\linewidth}{!}{
    		\begin{tabular}{ccccccc} 
    			\toprule
    			\mr{2}{\tabincell{c}{ \Th{Gallery}\\\Th{Model}}}&\mr{2}{\tabincell{c}{ \Th{Subspace}\\\Th{Number} $M$}} & \mr{2}{\tabincell{c}{\Th{Training}\\\Th{Set} $\mathcal{T}$}} &\mc{2}{\Th{Medium}} & \mc{2}{\Th{Hard}} \\ \cmidrule(l){4-5} \cmidrule(l){6-7}
    			& & &$\mathcal{R}$Oxf & $\mathcal{R}$Par & $\mathcal{R}$Oxf & $\mathcal{R}$Par \\ 
    			\midrule
    			\multirow{6}*{R101-GeM~\cite{radenovic2018fine}}& 2 & \mr{6}{SfM-120k}& 51.1  & 66.6  & 28.5 & 37.9 \\
    			& 4 & & 53.5 & 68.5 & 29.6 & 41.1 \\
    			& 8 & & 55.1 & 69.3 & 30.5 & 42.6 \\
    			& 16 & & 56.4 & 70.9 & 32.7 & 45.1 \\
    			& 32 & & 62.0 & 74.6 & 35.8 & 50.9 \\
    			& 64 & & \tb{63.9} & \tb{76.5} & \tb{37.9} & \tb{53.5} \\
    			\midrule
    			\multirow{6}*{R101-DELG~\cite{DELG}}& 2 & \mr{6}{GLDv2}& 68.4  & 80.8  & 46.4 & 64.4 \\
    			& 4 & & 70.2 & 83.4 & 49.8 & 68.2 \\
    			& 8 & & 73.9 & 87.0 & 54.8 & 72.7 \\
    			& 16 & & 77.6 & 87.2 & 59.2 & 73.8 \\
    			& 32 & & 77.8 & \tb{88.3} & \tb{60.0} & \tb{75.0} \\
    			& 64 & & \tb{78.2} & 88.0 & 59.8 & 74.7 \\
    			\bottomrule
    		\end{tabular}
            }
	\end{center}\vspace{-0.5em}
	\caption{mAP (asymmetric) \textbf{comparison of different number $M$ of subspaces}. $\tau_q$ and $\tau_g$ are set as 1.0 and 0.1, respectively. MobileNetV2~\cite{Mobilenetv2} is used as query model. R101-GeM and R101-DELG denote the ResNet101 trained by GeM~\cite{radenovic2018fine} and DELG~\cite{DELG}, respectively.\vspace{-1.0em}}
	\label{tab:M}
\end{table}

\subsection{Comparison with State-of-the-art Methods}
\noindent \tb{mAP Comparison}. In Table~\ref{tab:state-of-the-art}, we provide a comprehensive comparison of our proposed approach with state-of-the-art methods on various benchmark datasets. To evaluate the effectiveness of our method under different scenarios, we conduct experiments using different query models, gallery models, and training datasets. We compare the performance of two lightweight query models, MobileNetV2~\cite{Mobilenetv2} and EfficientNetB3~\cite{Efficientnet}, two large gallery models with varying performance, R101-GeM~\cite{radenovic2018fine} and R101-DELG~\cite{DELG}, and two training datasets with different sizes, SfM-120k~\cite{radenovic2018fine} and GLDv2~\cite{weyand2020google}. The first six rows in Table~\ref{tab:state-of-the-art} illustrate the performance of our method when using both large models (R101-GeM and R101-DELG) and small models (MobileNetV2 and EfficientNetB3) under a symmetrical setting.

We first evaluate our approach using R101-GeM as the gallery model and SfM-120k as the training set. Our method outperforms the most effective solution to \emph{asymmetric image retrieval} in AML~\cite{AML}, \ie, direct feature regression (Reg), by a large margin. Furthermore, our approach achieves consistently superior or comparable performance compared to CSD~\cite{CSD}, which takes neighbor similarity into consideration. For instance, when MobileNetV2 is deployed as the query model, our method still outperforms CSD in most settings, with an mAP improvement of $1.59\%$ on the $\gR$Par + $\gR$1M dataset. It is worth mentioning that in CSD, retrieved real data points are used to calculate neighbor similarity, whereas our method generates a significant number of anchor points in the embedding space of the gallery model, enabling a more detailed characterization of the spatial structure than that obtained using real data points.

Next, we evaluate our method with R101-DELG as the gallery model and GLDv2 as the training set. Our approach achieves better performance than the best previous method in most cases, regardless of whether the query model is MobileNetV2 or EfficientNetB3. \eg, when MobileNetV2 is deployed as the query model, our method outperforms CSD by $2.94\%$ and $2.54\%$ on the $\gR$Oxf dataset with Medium and Hard protocols, respectively. Similarly, on the $\gR$Par dataset with Medium and Hard protocols, our method outperforms CSD by $1.07\%$ and $1.63\%$, respectively. All these results convincingly demonstrate the superiority of our approach.

While we acknowledge that our approach may perform less favorably than symmetric retrieval when large models are deployed on both the query and gallery sides, it is important to highlight the efficiency of our approach when a smaller model is deployed on the query side. With only $5.8\%$ of the computational FLOPS required by a model like ResNet101~\cite{resnet}, our approach achieves a remarkable $90\%$ performance. Additionally, it is clear from Table~\ref{tab:state-of-the-art} that symmetric retrieval performance suffers when small models are deployed on both sides of the query and the database, and our asymmetric retrieval approach better balances performance and computational complexity in this case.

\noindent \tb{Discussion about Training Overhead}. Both our method and CSD introduce additional time overhead during training. As for CSD~\cite{CSD}, it performs retrieval in additional databases to obtain nearest neighbors during each iteration of the training process. In the CSD paper, the authors take the training set as an additional database. When dealing with GLDv2~\cite{weyand2020google} with $1,264,376$ images, the time overhead of this extra retrieval step becomes non-negligible. For instance, a single retrieval operation in this scenario incurs a latency of 0.105 seconds, and training 5 epochs with multi-threading acceleration takes approximately 24 hours.

In contrast, our approach introduces online computation of structural similarity, which indeed adds some time overhead. Specifically, our method divides features into 64 subvectors, each quantized to 256 clustering centers. The time overhead for computing structural similarity is approximately 16 ms. Note that even with this extra time overhead, based on the structural similarity, our approach costs much less training time than CSD with linear retrieval step. In detail, the total time overhead for training 5 epochs amounts to about 11 hours. Therefore, our method achieves better retrieval performance and costs less training time than CSD.

\begin{figure}[htb]
	\centering
        \hspace{-30pt}
	\resizebox{0.95\linewidth}{!}{
		\begin{subfigure}[b]{0.5\linewidth}  
			\centering 
			\input{plots/dim_plot_gldv2_private}
		\end{subfigure}
		\hfill
		\hspace{20pt}
		\begin{subfigure}[b]{0.5\linewidth}   
			\centering 
			\input{plots/map_vs_flops}
		\end{subfigure}
		
	}
	\caption{Comparison of the \textbf{time overhead} and \textbf{storage complexity} of different anchor generation methods. ``$k$-means'' means a flatten $k$-means quantizer and ``spectral'' denotes for spectral clustering. PQ$_{32\|256}$ denotes that we split the feature vector into 32 subvectors, with each subvector quantized to $256$ centroids.\vspace{-1.0em}}
	\label{fig:ablation_time}
\end{figure}
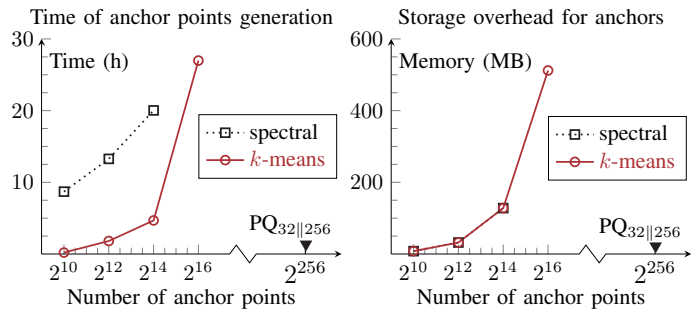

\subsection{Ablation Study}
\noindent \tb{Comparison with different clustering options}. Our method generates anchor points in the embedding space of the gallery model, whose number is related to the granularity of the space division. In this experiment, we compare product quantization with the flat $k$-means and spectral clustering~\cite{spectral_1,spectral_2}.

As shown in Table~\ref{tab:Flatten_Kmeans}, adopting flatten $k$-means or spectral clustering leads to severe performance degradation, which is mainly due to the coarse granularity of the space partition. The performance gradually increases as the number of anchor images increases, which shows the need for a large number of anchor points. However, as show in Figure~\ref{fig:ablation_time}, when the number of required centroids is large, $k$-means and spectral clustering lead to heavy computations and unaffordable time overheads, making it difficult to further scale up the number of anchor points. This limits the granularity of the partitioning for the space, which makes the relationship between features and anchor points fail to reflect the structure of the space well. Besides, when adopting spectral and $k$-means clustering, we need to save a large number of centroid vectors. Thus, product quantization is more suitable.

\noindent \tb{Number of Subspaces}. Table~\ref{tab:M}
shows the mAP of our method with different numbers $M$ of subspaces. As $M$ increases, the performance increases at all settings. When $M$ is small, the number of anchor points is small. The division granularity of the gallery embedding space is too coarse to delicately characterize its structure. In contrast, when $M$ is large, the number of equivalent anchor points is large, \eg, when $M$ is $32$, it reaches $256^{32}$. By constraining the consistency of structure similarities between feature embeddings of the same training sample, the embedding spaces of the query and gallery models are well aligned.

\begin{table}[t]
	\begin{center}
		\small
		\setlength{\extrarowheight}{0.0pt}
		\setlength{\tabcolsep}{2.0pt}
            \resizebox{\linewidth}{!}{
    		\begin{tabular}{ccccccc} 
    			\toprule
    			\mr{2}{\tabincell{c}{ \Th{Gallery}\\\Th{Model}}}&\mr{2}{\tabincell{c}{ \Th{Similarity}\\\Th{Type}}} & \mr{2}{\tabincell{c}{\Th{Training}\\\Th{Set} $\mathcal{T}$}} &\mc{2}{\Th{Medium}} & \mc{2}{\Th{Hard}} \\ \cmidrule(l){4-5} \cmidrule(l){6-7}
    			& & & $\mathcal{R}$Oxf & $\mathcal{R}$Par & $\mathcal{R}$Oxf & $\mathcal{R}$Par \\ 
    			\midrule
    			\multirow{2}*{R101-GeM~\cite{radenovic2018fine}}& Equation~(\ref{equ:L_2}) & \mr{2}{SfM-120k} & 56.4  & 69.7  & 32.1 & 44.0 \\
    			& Equation~(\ref{equ:cosine}) & & \tb{63.9} & \tb{76.5} & \tb{37.9} & \tb{53.5} \\
    			\midrule
    			\multirow{2}*{R101-DELG~\cite{DELG}}& Equation~(\ref{equ:L_2}) & \mr{2}{GLDv2}& 65.4  & 79.6  & 42.9 & 59.9 \\
    			& Equation~(\ref{equ:cosine}) & & \tb{77.8} & \tb{88.3} & \tb{60.0} & \tb{75.0} \\
    			\bottomrule
    		\end{tabular}
            }
	\end{center}
	\caption{mAP (asymmetric) comparison of \textbf{different similarity types}. MobileNetV2~\cite{Mobilenetv2} is used as query model. R101-GeM and R101-DELG denote the ResNet101 trained by GeM and DELG, respectively.}
	\label{tab:similarity_type}
\end{table}

\begin{table}[t]
	\begin{center}
		\small
		\setlength{\extrarowheight}{0.0pt}
		\setlength{\tabcolsep}{2.0pt}
            \resizebox{\linewidth}{!}{
    		\begin{tabular}{ccccccc} 
    			\toprule
    			\mr{2}{\tabincell{c}{ \Th{Gallery}\\\Th{Model}}} & \mr{2}{\tabincell{c}{\Th{Temperature}\\ $\tau_{g}$}} & \mr{2}{\tabincell{c}{\Th{Training}\\\Th{Set} $\mathcal{T}$}}&\mc{2}{\Th{Medium}} & \mc{2}{\Th{Hard}} \\ \cmidrule(l){4-5} \cmidrule(l){6-7}
    			& & &$\mathcal{R}$Oxf & $\mathcal{R}$Par & $\mathcal{R}$Oxf & $\mathcal{R}$Par \\ 
    			\midrule
    			\multirow{5}*{R101-GeM~\cite{radenovic2018fine}}&0.00& \mr{5}{SfM-120k}&62.7 & 73.8 & 35.8 & 49.0 \\
    			&0.01 & &\tb{65.0} & 74.5 & \tb{38.9} & 50.7 \\
    			&0.1 & &63.9 & \tb{76.5} & 37.9 & \tb{53.5} \\
    			&0.2 & &63.5  & 74.6  & 36.8 & 49.7 \\
    			&0.5 & &58.7  & 71.4  & 31.8 & 46.0 \\
    			\midrule
    			\multirow{5}*{R101-DELG~\cite{DELG}}&0.00& \mr{5}{GLDv2}&72.6 & 86.1 & 53.2 & 71.4 \\
    			&0.01 & &75.6 & 86.9 & 57.4 & 73.3 \\
    			&0.1 & &\tb{77.8}  & \tb{88.3}  & \tb{60.0} & \tb{75.0} \\
    			&0.2 & &75.0  & 87.5  & 55.1 & 73.9 \\
    			&0.5 & &62.8  & 79.3 & 45.6 & 60.7 \\
    			\bottomrule
    		\end{tabular}
            }
	\end{center}
	\caption{Analysis about the \textbf{temperature $\tau_g$}. MobileNetV2~\cite{Mobilenetv2} is used as query model and  $\tau_q$ is set to 1.0. R101-GeM and R101-DELG denote the ResNet101 trained by GeM and DELG, respectively.}\vspace{-1.0em}
	\label{tab:temperature}
\end{table}

\noindent \tb{Similarity Type}. As shown in Table~\ref{tab:similarity_type}, we explore two types of similarities, including negative Euclidean distance and Cosine similarity. When negative Euclidean distance is adopted as the similarity strategy, $s(\cdot,\cdot)$ is formulated as:
\begin{eqnarray}\label{equ:L_2}
	\begin{aligned}
		s(u_i(\bm{g}),\bm{C}_{i,:}^j) = -\| \bm{C}_{i,:}^j - u_i(\bm{g})\|_2.
	\end{aligned}
\end{eqnarray}
``Cosine similarity'' leads to better performance.  The negative Euclidean distance ranges from 0 to $-\infty$, and the probabilities $\bm{p}_{i,:}^{g}$ and $\bm{p}_{i,:}^{q}$ obtained after the softmax function do not reflect well the relationship between the feature vectors and the anchor points. On the contrary, the cosine similarity ranges from -1 to 1, which makes the final probability distribution more discriminative.

\noindent \tb{Soft \emph{vs}\onedot Hard Assignment}. In Table~\ref{tab:temperature}, we demonstrate the effect of temperature $\tau_g$ in Equation~(\ref{equ:probalility}). The results for the hard assignment case are denoted by $\tau_g = 0.00$. Choosing a small $\tau_g$, which makes the probability $p_i^{g}$ sharper (closer to hard assignment), leads to better performance. However, in the extreme case of hard assignment, the performance decreases. Hard assignment ignores the relationship between features and anchor points, which characterizes the space structure and contains more useful knowledge. 

\begin{figure}[htb]
	\centering
        \hspace{-30pt}
	\resizebox{0.95\linewidth}{!}{
		\begin{subfigure}[b]{0.5\linewidth}  
			\centering 
			\input{plots/dim_plot_par_hard_}
		\end{subfigure}
		\hfill
		\hspace{17pt}
		\begin{subfigure}[b]{0.5\linewidth}   
			\centering 
			\input{plots/dim_plot_par_medium_}
		\end{subfigure}
		
	}
	\caption{\textbf{Comparison of mAP (asymmetric retrieval) of different methods to generate anchor points}.
		The mAP is the average of two difficulty settings, medium (left) and high (right).
		R101-DELG~\cite{DELG} and MobileNetV2~\cite{Mobilenetv2} are deployed as the gallery and query models, respectively. ``random'' denotes for random selecting image feautres as anchor points. ``$k$-means'' means a flatten $k$-means quantizer and ``spectral'' refers to spectral clustering of the data. PQ$_{32\|256}$ denotes that we split the feature vector into 32 subvectors, with each subvector quantized to $256$ centroids.}
	\label{fig:ablation_output_dim}
\end{figure}
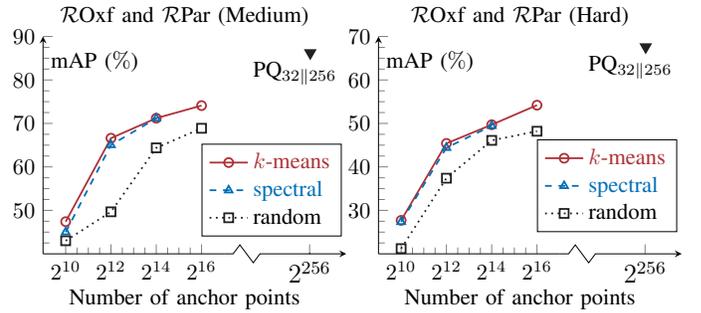

\noindent \tb{Scalability}. The structure similarity generated by the gallery model serves as a pseudo-label to supervise the learning of the query model. While the gallery model can be trained using labeled data if available, the query model is exclusively trained through pseudo-labels generated by the gallery model. This means that it is feasible to leverage the vast amount of unlabeled data during the training phase of the query model. In Table~\ref{tab:Train_dataset_AUG}, we divide the GLDv2 dataset~\cite{weyand2020google} into 10 random splits and train the query models using different amounts of data. In both settings, \eg, using R101-GeM and R101-DELG as the gallery models, the performance gradually improves as the number of training data increases. Our approach does not use any annotations of the training data but only exploits the knowledge provided by the gallery model. Thus, it is possible for our method to improve the performance of the query model using a large amount of unlabeled data. 

We further validate this in Table~\ref{tab:Train_dataset_OOD}, which shows that convincing results are achieved when ImageNet~\cite{deng2009imagenet} is adopted as the training set. Our approach does not directly regress features but uses the relationships between features and anchor points as knowledge, which somewhat weakens the effect of image data distribution bias. Notably, we achieve remarkable performance when we adopt the distractor set $\mathcal{R}$1M as the training set, which further illustrates that our approach is able to utilize the available unlabeled data to train query models.

\noindent \tb{Different methods to generate anchor points}. In this section, we explore different ways to generate anchor points such as random selection from gallery features and spectral clustering. As shown in Figure~\ref{fig:ablation_output_dim}, randomly selecting anchor points yields the least favorable results. This observed performance drop may be attributed to the fact that randomly chosen anchor points do not reflect the density of the data distribution, making it difficult for the structural similarity to accurately portray the structural information in the gallery space.

\begin{table}[t]
	\begin{center}
		\small
		\setlength{\extrarowheight}{0.0pt}
		\setlength{\tabcolsep}{2pt}
            \resizebox{\linewidth}{!}{
    		\begin{tabular}{ccccccc} 
    			\toprule
    			\mr{2}{ \tabincell{c}{ \Th{Gallery}\\\Th{Model}}}&\mr{2}{\tabincell{c}{\Th{Training}\\\Th{Set} $\mathcal{T}$}} & \mr{2}{\tabincell{c}{\Th{Image}\\\Th{Numbers}}}& \mc{2}{\Th{Medium}} & \mc{2}{\Th{Hard}} \\ \cmidrule(l){4-5} \cmidrule(l){6-7}
    			& & & $\mathcal{R}$Oxf & $\mathcal{R}$Par & $\mathcal{R}$Oxf & $\mathcal{R}$Par \\ \midrule
    			\multirow{4}*{R101-GeM~\cite{radenovic2018fine}}
    			& GLDv2 ($\times 0.1$) & 128,078 & 63.5  & 75.6  & 37.2 & 53.6 \\
    			& GLDv2 ($\times 0.2$) & 256,156 & 64.1  & 76.8  & 38.5 & 54.3 \\
    			& GLDv2 ($\times 0.3$) & 384,234 & 64.2  & \tb{77.6}  & 39.2 &
    			\tb{56.7} \\
    			& GLDv2 ($\times 0.4$) & 512,312 & \tb{64.9}  & 77.3  & \tb{40.1} &
    			56.2 \\
    			
    			\midrule
    			
    			\multirow{4}*{R101-DELG~\cite{DELG}}
    			& GLDv2 ($\times 0.1$) & 128,078 & 75.0  & 86.3  & 55.1 & 71.9 \\
    			& GLDv2 ($\times 0.2$) & 256,156 & 76.7  & 87.0  & 58.1 & 72.9 \\
    			& GLDv2 ($\times 0.3$) & 384,234 & \tb{77.1}  & 87.3  & 59.3 & 73.4 \\
    			& GLDv2 ($\times 0.4$) & 512,312 & 77.0  & \tb{87.6}  & \tb{59.4} & \tb{74.1} \\
    			\bottomrule
    		\end{tabular}
            }
	\end{center}
	\caption{mAP (\emph{asymmetric}) comparison of \textbf{different dataset size}. ($\times x$) denotes the small dataset formed by randomly selecting $x$ proportion of images from the full GLDv2 dataset~\cite{weyand2020google}. MobileNetV2~\cite{Mobilenetv2} is used as query model. R101-GeM and R101-DELG denote the ResNet101 trained by GeM and DELG, respectively.}
	\label{tab:Train_dataset_AUG}
\end{table}

\begin{table}[t]
	\begin{center}
		\small
		\setlength{\extrarowheight}{0.0pt}
		\setlength{\tabcolsep}{2.0pt}
            \resizebox{\linewidth}{!}{
    		\begin{tabular}{ccccccc} 
    			\toprule
    			\mr{2}{ \tabincell{c}{ \Th{Gallery}\\\Th{Model}}}&\mr{2}{\tabincell{c}{\Th{Training}\\\Th{Set} $\mathcal{T}$}} & \mr{2}{\tabincell{c}{\Th{Image}\\\Th{Numbers}}}& \mc{2}{\Th{Medium}} & \mc{2}{\Th{Hard}} \\ \cmidrule(l){4-5} \cmidrule(l){6-7}
    			& & & $\mathcal{R}$Oxf & $\mathcal{R}$Par & $\mathcal{R}$Oxf & $\mathcal{R}$Par \\ \midrule
    			\multirow{4}*{R101-GeM~\cite{radenovic2018fine}}&SfM-120k & 91,642 & 63.9 & 76.5 & 37.9 & 53.5 \\
    			&GLDv2 & 1,280,787 & \tb{65.2}  & \tb{77.5}  & \tb{40.1} & \tb{56.8} \\
    			& $\mathcal{R}$1M &  1,001,001 & 64.4  & 76.8  & 39.6 & 55.6 \\
    			& ImageNet & 1,281,167 & 57.8 & 74.4 & 31.9 & 52.6 \\
    			
    			\midrule
    			
    			\multirow{4}*{R101-DELG~\cite{DELG}}
    			&SfM-120k & 91,642 & 75.4  & 84.4  & 55.1 & 68.3 \\
    			& GLDv2 & 1,280,787 & \tb{77.8}  & \tb{88.3}  & \tb{60.0} & \tb{75.0} \\
    			& $\mathcal{R}$1M &  1,001,001 & 75.5  & 86.1  & 56.9 & 72.5 \\
    			& ImageNet  & 1,281,167 & 56.6  & 76.4  & 38.3 & 59.0 \\
    			\bottomrule
    		\end{tabular}
            }
	\end{center}
	\caption{mAP (\emph{asymmetric}) comparison of \textbf{different training datasets}. MobileNetV2~\cite{Mobilenetv2} is used as query model. R101-GeM and R101-DELG denote the ResNet101 trained by GeM and DELG, respectively.}
	\label{tab:Train_dataset_OOD}
\end{table}

\begin{figure*}[htb]
	\begin{center}
		\resizebox{1.0\linewidth}{!}{
			\includegraphics[width=\linewidth]{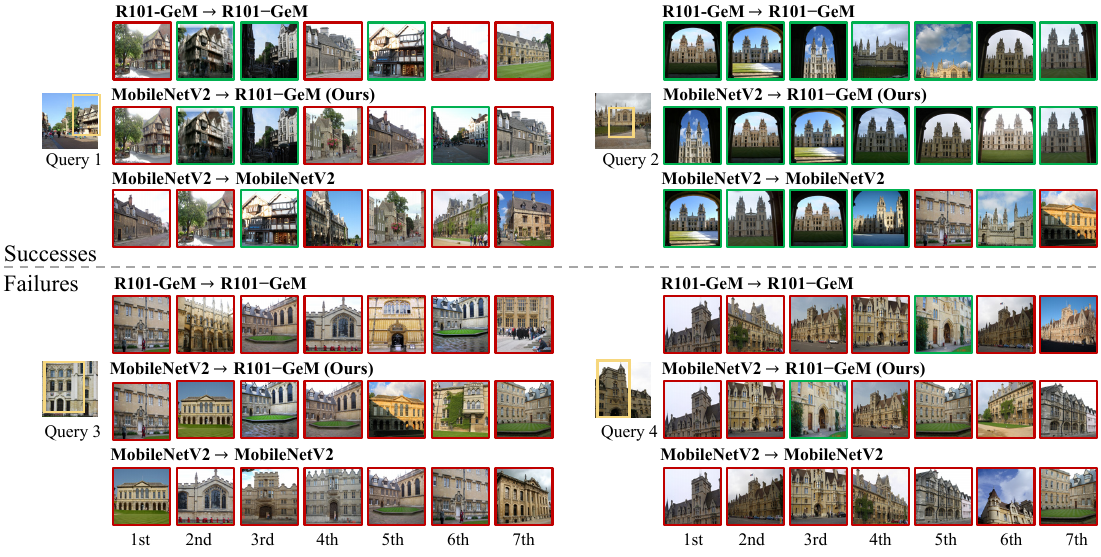}%
		}
	\end{center}
	\vspace*{-1.0em}
	\caption{\tb{Some successes and failures of our approach on $\mathcal{R}$Oxford5k}. In each group, the notation format \textbf{``query model $\rightarrow$ gallery model''} on the top means embedding queries with the query model and retrieving in a gallery set embedded by the gallery model. The image with the {\color{orange}{orange}} border on the left is a query, the first row shows the results of symmetric retrieval when the large gallery model deployed on both query and gallery sides. The second row shows the results of our method under asymmetric setting. The third row shows the results of symmetric retrieval when the lightweight query model is deployed on both sides. Images with {\color{green}{green}} borders are true positive images and images with {\color{red}{red}} borders are false positive images.}
	\label{fig:retrieval_results}
\end{figure*}

\begin{figure}[htb]
	\centering
	\resizebox{\linewidth}{!}{
		\input{plots/lightweight_query_models}
	}
	\caption{ Analysis of \textbf{different model variants}. We use R101-DELG~\cite{DELG} as gallery model $\phi_g(\cdot)$ and compare different architectures as query models $\phi_q(\cdot)$. \emph{Symmetric}: Query and gallery images are both embedded by $\phi_q(\cdot)$; \emph{Asymmetric}: Query and gallery images are embedded by $\phi_q(\cdot)$ and $\phi_g(\cdot)$, respectively. EB0: EfficientNetB0~\cite{Efficientnet}; EB1: EfficientNetB1~\cite{Efficientnet}; EB2: EfficientNetB2~\cite{Efficientnet}; SV2: ShuffleNetV2~\cite{shufflenetv2}; SV2$^{0.5\times}$: ShuffleNetV2 (0.5 $\times$)~\cite{shufflenetv2}.}
	\label{fig:Model}
\end{figure}
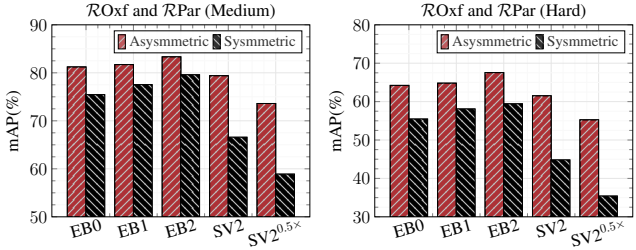

\begin{figure}[htb]
	\centering
	\begin{subfigure}{0.49\linewidth}
		\includegraphics[width=1.0\linewidth]{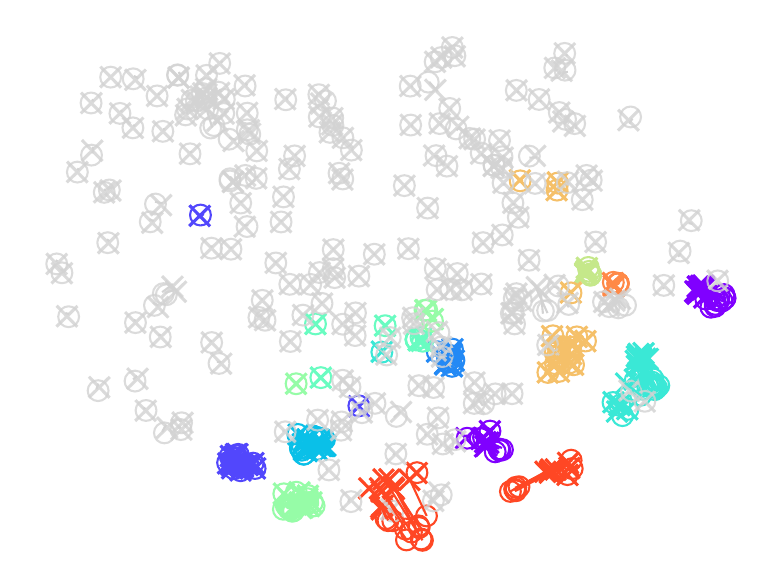}
		\caption{$\mathcal{R}$Oxf}
		\label{fig:short-a}
	\end{subfigure}
	\hfill
	\begin{subfigure}{0.49\linewidth}
		\includegraphics[width=1.0\linewidth]{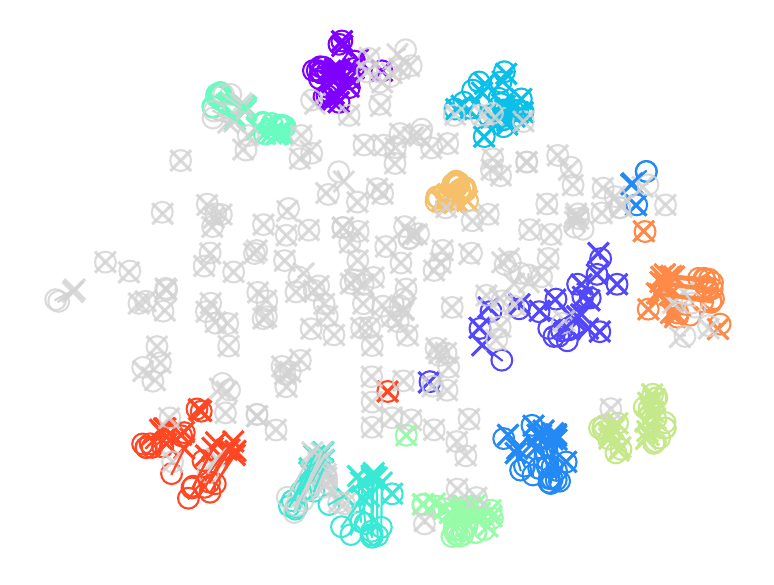}
		\caption{$\mathcal{R}$Par}
		\label{fig:short-b}
	\end{subfigure}
	\caption{\textbf{T-SNE embeddings of $\gR$Oxf and $\gR$Par datasets}. MobileNetV2~\cite{Mobilenetv2} and R101-GeM~\cite{radenovic2018fine} are used as query and gallery models. Different colors represent different buildings and gray denotes distractor images. We randomly select $10$ samples for each building category and $100$ in distractors. $\circ$ and $\times$ denote gallery and query models, respectively. A line connects the two representatives of each example.}
	\label{fig:tsne}
\end{figure}

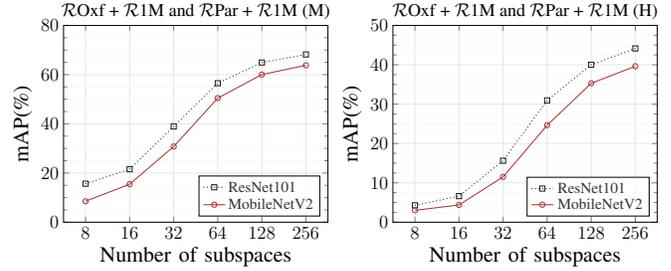
\begin{figure}[tb]
	\centering
	\resizebox{\linewidth}{!}{
		\input{plots/PQ}
	}
	\caption{mAP comparison when \textbf{PQ~\cite{PQ} is used during asymmetric and symmetric retrieval}. Horizontal axis denotes that we split the feature into $M$ subvectors. The number of centroids per subvector is set to 256. R101-DELG~\cite{DELG} is deployed as the gallery model. We report the average performance of our method on $\gR$Oxf + $\gR$1M and $\gR$Par + $\gR$1M datasets with Medium (M) and Hard (H) protocols~\cite{radenovic2018revisiting}.}
	\label{fig:PQ_results}
\end{figure}

\begin{table}[htb]
	\begin{center}
		\small
		\setlength{\extrarowheight}{0.0pt}
		\setlength{\tabcolsep}{4.0pt}
            \resizebox{\linewidth}{!}{
    		\begin{tabular}{lccc} 
    			\toprule
    			\multirow{2}{*}{~~\Th{Method}} & \mr{2}{\tabincell{l}{\Th{\ Retrieval}\\\Th{latency (ms)}}}  & \mc{2}{\Th{Memory} (\Th{MB})} \\ \cmidrule(l){3-3} \cmidrule(l){4-4}
    			& & $\mathcal{R}$Oxf + $\mathcal{R}$1M & $\mathcal{R}$Par + $\mathcal{R}$1M \\ \midrule
    			~~PQ$_{8\|256}$   & $29.69$  & $7.68$    & $7.69$      \\
    			~~PQ$_{16\|256}$  & $30.57$  & $15.35$   & $15.37$     \\
    		    ~~PQ$_{32\|256}$  & $32.97$  & $30.70$   & $30.75$     \\
    			~~PQ$_{64\|256}$  & $34.26$  & $61.40$   & $61.50$     \\
    			~~PQ$_{128\|256}$ & $41.15$  & $122.80$  & $122.90$    \\
    			~~PQ$_{256\|256}$ & $51.87$  & $245.60$  & $245.98$    \\
    			No quantization & $152.12$  & $7,782.40$  & $7,783.70$    \\
    			\bottomrule
    		\end{tabular}
            }
	\end{center}
	\caption{\textbf{Time and memory cost}. We report average search time on a
		single thread CPU (Intel Xeon CPU E5-2640 v4 @ 2.40GHz) and memory consumption for the gallery sets with 1M distractor images. }
	\label{tab:speed_memory}
\end{table}

\noindent \tb{Various Lightweight Models}. In this section, we experiment with more lightweight models, whose computational complexity (in FLOPS) is shown in Table~\ref{tab:net_params}, as query models $\phi_q(\cdot)$.
In Figure~\ref{fig:Model}, \emph{symmetric} means that the query and gallery images are both processed using $\phi_q(\cdot)$, while \emph{asymmetric} means that the query and gallery images are processed using $\phi_q(\cdot)$ and $\phi_g(\cdot)$, respectively. The performance becomes better as the model parameters and \Th{Flops} increase under both \emph{asymmetric} and \emph{symmetric} settings. Notably, the performance improvement of the \emph{asymmetric} setting over the \emph{symmetric} setting is more obvious when the number of model parameters is small, \eg, ShuffeNetV2 (0.5$\times$), which indicates the advantage of \emph{asymmetric image retrieval} in resource-constrained scenarios. In practical scenarios, it needs to compromise the computational complexity and retrieval accuracy to select an appropriate query model.

\noindent \tb{Visualization of Retrieval Results}. In Figure~\ref{fig:retrieval_results}, we show some examples of the success and failure of our approach. According to the retrieval results of queries 1 and 2, deploying large gallery models on both the query and gallery sides results in the highest retrieval performance. However, it is important to acknowledge that in resource-constrained scenarios, such as on mobile devices, deploying large models may not be feasible. When lightweight models are used on both sides, there is a noticeable degradation in retrieval performance. Our approach takes an asymmetric approach, where lightweight, smaller models are deployed on the query side while high-performing large models are used on the gallery side. Additionally, we train the query model to be compatible with the gallery model, striking a balance between computational complexity and retrieval performance. 
		
However, it is worth noting that there are instances where retrieval results remain sub-optimal even when large gallery models are deployed on both sides. These cases represent challenges that our approach, or any method, may face. The small model trained by our method is constrained to maintain structural similarity with the gallery model, and there are scenarios where it may struggle with certain query images.

\noindent \tb{Qualitative Results}. Figure~\ref{fig:tsne} shows the embeddings of some $\gR$Oxf and $\gR$Par images, each processed by a gallery and a query model. For \emph{asymmetric image retrieval}, it is crucial to keep the feature compatibility between query and model models. During training, anchor points are shared by both query and gallery models. We restrict the similarities between two features of the same training sample and anchor points to be consistent, which keeps the structure similarity.

\noindent \tb{Memory \emph{vs}\onedot Search Accuracy}. In Figure~\ref{fig:PQ_results}, we adopt PQ during the online retrieval, with the corresponding retrieval latency and memory consumption shown in Table~\ref{tab:speed_memory}. PQ is parametrized by the number of sub-vectors $M$ and the number of quantizers per sub-vector $K$, producing a code of length $M \times \text{log}_2K$. As $M$ increases, the accuracy of retrieval gradually approximates the direct feature comparison. When $M=256$, quantization saves $96.8\%$ of memory and $65.9\%$ of retrieval latency with slight performance degradation. In real-word applications, we choose the appropriate $M$ to achieve the performance-memory trade-off.

\section{Conclusion}
In this paper, we propose a structure similarity preserving approach to achieve feature consistency between query and gallery models for asymmetric retrieval. First, we employ product quantization to generate a large number of anchor points in the embedding space of the gallery to characterize its space structure. Then, these anchor points are shared between query and gallery models. The relationships between each training sample and anchor points are considered as structure similarity and constrained to be consistent across different models. This allows the query model to focus less on the feature ``details'' of the gallery model and more on the overall space structure. Besides, our method does not utilize any annotation from training set, and it is possible for the proposed method to utilize large-scale unlabeled training data, even from different domains. This shows the generalizability of our approach. Extensive experiments show that our method achieves better performance than state-of-the-art asymmetric retrieval methods.

\bibliography{TMM}
\bibliographystyle{IEEEtran}

\begin{IEEEbiography}[{\includegraphics[width=1in,height=1.25in,clip,keepaspectratio]{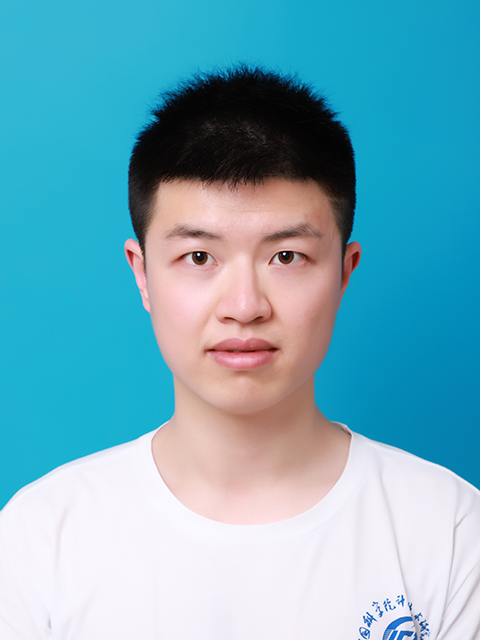}}]{Hui Wu} is currently pursuing the Ph.D. degree in information and communication engineering with the Department of Data Science, from the University of Science and Technology of China. 

His research interests include image retrieval, multimedia information retrieval and computer vision.
\end{IEEEbiography}

\begin{IEEEbiography}[{\includegraphics[width=1in,height=1.25in,clip,keepaspectratio]{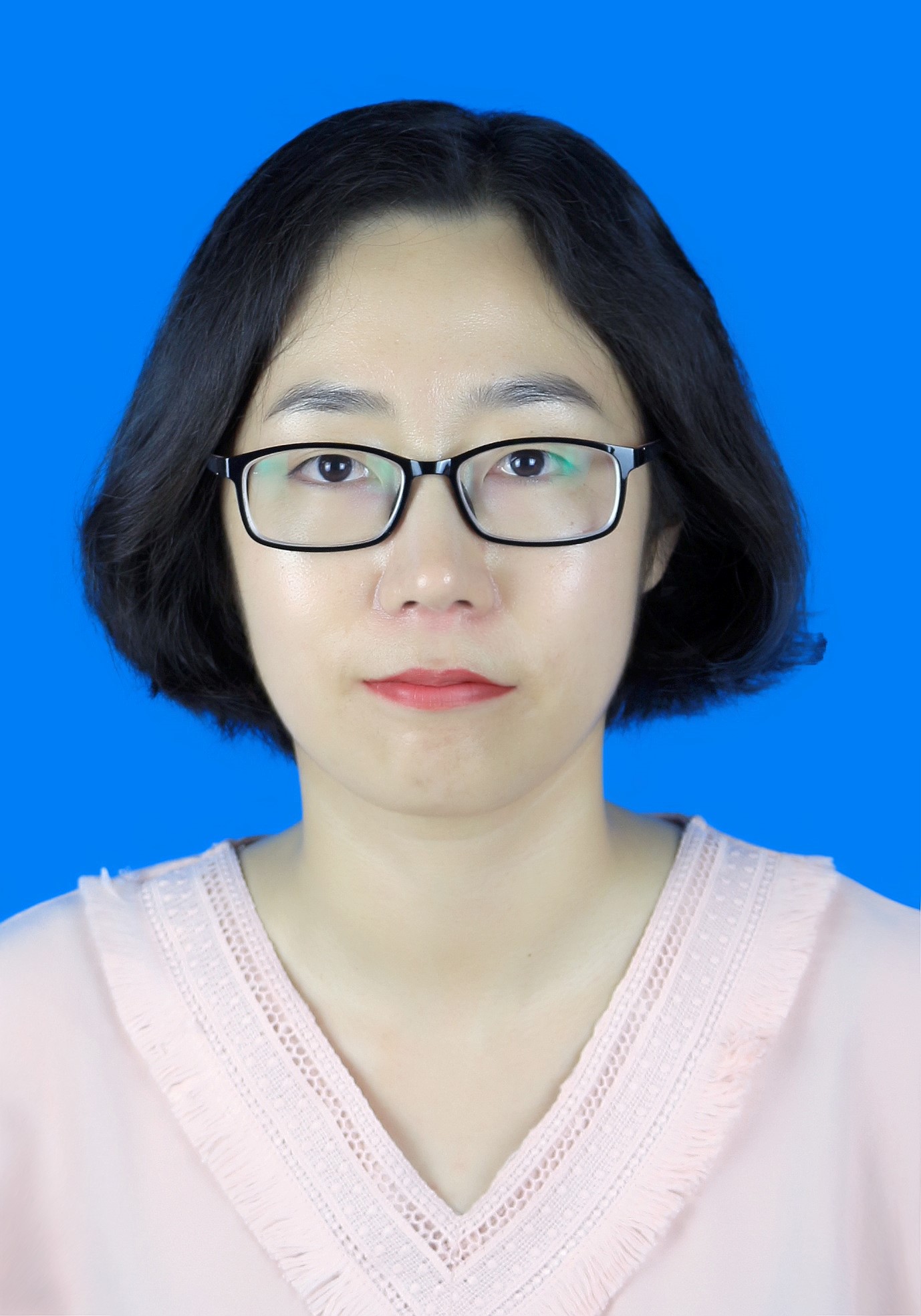}}]{Min Wang} received the B.E., and Ph.D degrees in electronic information engineering from University of Science and Technology of China (USTC), in 2014 and 2019, respectively. She is working in Institute of Artificial Intelligence, Hefei Comprehensive National Science Center. 
	
Her current research interests include binary hashing, multimedia information retrieval and computer vision.
\end{IEEEbiography}

\begin{IEEEbiography}[{\includegraphics[width=1in,height=1.25in,clip,keepaspectratio]{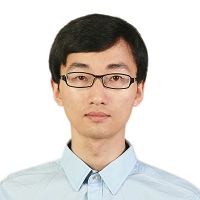}}]{Wengang Zhou} (S'20) received the B.E. degree in electronic information engineering from Wuhan University, China, in 2006, and the Ph.D. degree in electronic engineering and information science from the University of Science and Technology of China (USTC), China, in 2011. From September 2011 to September 2013, he worked as a postdoc researcher in Computer Science Department at the University of Texas at San Antonio. He is currently a Professor at the EEIS Department, USTC. 
	
	His research interests include multimedia information retrieval, computer vision, and computer game. In those fields, he has published over 100 papers in IEEE/ACM Transactions and CCF Tier-A International Conferences. He is the winner of National Science Funds of China (NSFC) for Excellent Young Scientists. He is the recepient of the Best Paper Award for ICIMCS 2012. He received the award for the Excellent Ph.D Supervisor of Chinese Society of Image and Graphics (CSIG) in 2021, and the award for the Excellent Ph.D Supervisor of Chinese Academy of Sciences (CAS) in 2022. He won the First Class Wu-Wenjun Award for Progress in Artificial Intelligence Technology in 2021. He served as the publication chair of IEEE ICME 2021 and won 2021 ICME Outstanding Service Award. He is currently an Associate Editor and a Lead Guest Editor of IEEE Transactions on Multimeida. 
\end{IEEEbiography}

\begin{IEEEbiography}[{\includegraphics[width=1in,height=1.25in,clip,keepaspectratio]{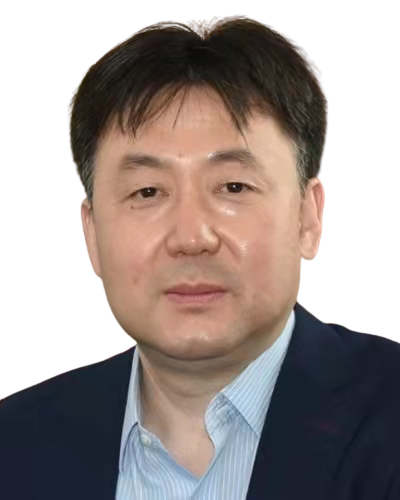}}]{Houqiang Li} (S'12, F'21) received the B.S., M.Eng., and Ph.D. degrees in electronic engineering from the University of Science and Technology of China, Hefei, China, in 1992, 1997, and 2000, respectively, where he is currently a Professor with the Department of Electronic Engineering and Information Science. 
	
	His research interests include image/video coding, image/video analysis, computer vision, reinforcement learning, etc.. He has authored and co-authored over 200 papers in journals and conferences. He is the winner of National Science Funds (NSFC) for Distinguished Young Scientists, the Distinguished Professor of Changjiang Scholars Program of China, and the Leading Scientist of Ten Thousand Talent Program of China. He is the associate editor (AE) of IEEE TMM, and served as the AE of IEEE TCSVT from 2010 to 2013. He served as the General Co-Chair of ICME 2021 and the TPC Co-Chair of VCIP 2010. He received the second class award of China National Award for Technological Invention in 2019, the second class award of China National Award for Natural Sciences in 2015, and the first class prize of Science and Technology Award of Anhui Province in 2012. He received the award for the Excellent Ph.D Supervisor of Chinese Academy of Sciences (CAS) for four times from 2013 to 2016. He was the recipient of the Best Paper Award for VCIP 2012, the recipient of the Best Paper Award for ICIMCS 2012, and the recipient of the Best Paper Award for ACM MUM in 2011.
\end{IEEEbiography}
\end{document}

%% file: math_commands.tex
\usepackage{amsmath,amsfonts,bm}

\def\eqref#1{equation~\ref{#1}}

\def\1{\bm{1}}

\DeclareMathAlphabet{\mathsfit}{\encodingdefault}{\sfdefault}{m}{sl}
\SetMathAlphabet{\mathsfit}{bold}{\encodingdefault}{\sfdefault}{bx}{n}

\def\gM{{\mathcal{M}}}

\def\gQ{{\mathcal{Q}}}
\def\gR{{\mathcal{R}}}

\def\gT{{\mathcal{T}}}

\def\gX{{\mathcal{X}}}



%% file: plots/dim_plot_gldv2_private.tex
\pgfmathsetmacro{\teasermarkersize}{2.5}

\begin{tikzpicture}
    \begin{groupplot}[
        group style={
            group size=2 by 1,
            horizontal sep=0pt
        },
        axis x line=bottom,
        axis y line=middle, 
        longlines/.style={
            shorten >=-10mm,
            shorten <=-10mm
        },
        legend style={at={(axis cs:16.0, 10.5pt)}, anchor=south west, font=\normalsize, fill opacity=1.0},
        ]
        
        \pgfplotstableread{
            gflops csd mix 	
            10  0.2 8.7
            12  1.81 13.27 
            14  4.68 20.03 
            16  26.98 nan
        }{\map}
        
        \nextgroupplot[
        grid=none,
        height=5cm, 
        xlabel={Number of anchor points},
        ylabel={Time (h)},
        xmin=9, xmax=17,
        xtick={10,12,14,16},
        xticklabels={$2^{10}$,$2^{12}$, $2^{14}$, $2^{16}$},
        ymin=0, ymax=30,
        minor tick num=3,
        title={Time of anchor points generation},
        title style={xshift=0.8cm, yshift=-5pt,font=\normalsize},
        grid style={line width=.1pt, draw=gray!5},
        major grid style={line width=.2pt,draw=gray!20},
        tick label style={font=\normalsize},
        tick align=inside,
        label style={font=\normalsize},
        legend style={font=\normalsize},
        x label style={xshift=0.8cm, yshift=0.1cm},
        y label style={yshift=-0.1cm},
        x axis line style={-},
        ]
        
        \addplot[line width =2pt, thick, dotted, color=\csdcolor, mark=square]      table[x=gflops,  y=mix] \map; 
        \leg{spectral}
        
        \addplot[line width =2pt, oursoxf]      table[x=gflops,  y=csd] \map; 
        \leg{\color{Maroon}{$k$-means}} 
        
        \nextgroupplot[
        grid=none,
        tick label style={font=\large},
        tick align=inside,
        hide y axis,  
        axis x discontinuity=crunch,
        width=3.5cm, 
        height=5cm, 
        minor tick num=3,
        xmin=20, xmax=25,
        ymin=0, ymax=10,
        xtick={23.5},
        xticklabels={$2^{256}$},
        ]

        \coordinate(a) at(23.5,0.0);
        \coordinate(b) at(23.5,0.438);
        \draw[-,dashed] (a) edge (b);
        
        \node[isosceles triangle, fill, \csdcolor, scale=0.4,rotate=30,isosceles triangle apex angle=60,] at (axis cs: 23.5, 0.438) {}; 
        \node[] at (axis cs:  22.9, 1.3) {PQ$_{32\|256}$}; 
        
    \end{groupplot}
	
\end{tikzpicture}

  

%% file: plots/map_vs_flops.tex
\pgfmathsetmacro{\teasermarkersize}{2.5}

\begin{tikzpicture}
	\begin{groupplot}[
		group style={
			group size=2 by 1,
			horizontal sep=0pt
		},
		axis x line=bottom,
		axis y line=middle, 
		longlines/.style={
			shorten >=-10mm,
			shorten <=-10mm
		},
		legend style={at={(axis cs:16.0, 205pt)}, anchor=south west, font=\normalsize, fill opacity=1.0},
		]
		
		\pgfplotstableread{
			gflops csd mix 	
			10  8 8
			12  32 32 
			14  128 128 
			16  512 nan 
		}{\map}
		
		\nextgroupplot[
		grid=none,
		height=5cm, 
		xlabel={Number of anchor points},
		ylabel={Memory (MB)},
		xmin=9, xmax=17,
		xtick={10,12,14,16},
		xticklabels={$2^{10}$,$2^{12}$, $2^{14}$, $2^{16}$},
		ymin=0, ymax=600,
		minor tick num=3,
		title={Storage overhead for anchors},
		title style={xshift=0.8cm, yshift=-5pt,font=\normalsize},
		grid style={line width=.1pt, draw=gray!5},
		major grid style={line width=.2pt,draw=gray!20},
		tick label style={font=\normalsize},
		tick align=inside,
		label style={font=\normalsize},
		legend style={font=\normalsize},
		x label style={xshift=0.8cm,yshift=0.1cm},
		y label style={yshift=-0.1cm},
		x axis line style={-},
		]
		
		\addplot[line width =2pt, thick, dotted, color=\csdcolor, mark=square]      table[x=gflops,  y=mix] \map; 
		\leg{spectral}
		
		\addplot[line width =2pt, oursoxf]      table[x=gflops,  y=csd] \map; 
		\leg{\color{Maroon}{$k$-means}} 
		
		\nextgroupplot[
		grid=none,
		tick label style={font=\large},
		tick align=inside,
		hide y axis,  
		axis x discontinuity=crunch,
		width=3.5cm, 
		height=5cm, 
		minor tick num=3,
		xmin=20, xmax=25,
		ymin=0, ymax=600,
		xtick={23.5},
		xticklabels={$2^{256}$},
		]
		
		\addplot[color=black, mark=+,mark size=3.5pt, thick=1pt, only marks] coordinates {(46.60,64.405)};
		\node [below] at (axis cs: 42.9,65.25) {\textcolor{black}{\normalsize 64.40}};

		\coordinate(a) at(23.5,0.0);
		\coordinate(b) at(23.5,2.0);
		\draw[-,dashed] (a) edge (b);
		
		\node[isosceles triangle, fill, \csdcolor, scale=0.4,rotate=30,isosceles triangle apex angle=60,] at (axis cs: 23.5, 21.0) {}; 
		\node[] at (axis cs:  22.9, 70) {PQ$_{32\|256}$}; 
		
	\end{groupplot}
	
\end{tikzpicture}

%% file: plots/dim_plot_par_hard_.tex
\pgfmathsetmacro{\teasermarkersize}{2.5}

\begin{tikzpicture}
	\begin{groupplot}[
		group style={
			group size=2 by 1,
			horizontal sep=0pt
		},
		axis x line=bottom,
		axis y line=middle, 
		longlines/.style={
			shorten >=-10mm,
			shorten <=-10mm
		},
		legend style={at={(axis cs:16.0, 44.0pt)}, anchor=south west, font=\normalsize, fill opacity=1.0},
		]
		
		\pgfplotstableread{
			gflops mixoxf csdoxf spectral	
			10  47.4 43.01 45
			12  66.6 49.67 65.05
			14 71.2 64.38 71.15
			16 74.1 68.88 nan
		}{\map}
		
		\nextgroupplot[
		grid=none,
		height=5cm, 
		xlabel={Number of anchor points},
		ylabel={mAP ($\%$)},
		xmin=9, xmax=17,
		xtick={10,12,14,16},
		xticklabels={$2^{10}$,$2^{12}$, $2^{14}$, $2^{16}$},
		ymin=40, ymax=90,
		minor tick num=3,
		title={$\gR$Oxf and $\gR$Par (Medium)},
		title style={xshift=0.8cm, yshift=-5pt,font=\normalsize},
		grid style={line width=.1pt, draw=gray!5},
		major grid style={line width=.2pt,draw=gray!20},
		tick label style={font=\normalsize},
		tick align=inside,
		label style={font=\normalsize},
		legend style={font=\normalsize},
		x label style={xshift=0.8cm, yshift=0.1cm},
		y label style={yshift=-0.1cm},
		x axis line style={-},
		]
		
		\addplot[line width =2pt, oursoxf]      table[x=gflops,  y=mixoxf] \map; 
		\leg{\color{Maroon}{$k$-means}} 

            \addplot[line width =2pt, oursoxfs]      table[x=gflops,  y=spectral] \map; 
		\leg{\color{\ourscolors}{spectral}}
		
		\addplot[line width =2pt, thick, dotted, color=\csdcolor, mark=square]      table[x=gflops,  y=csdoxf] \map;
		\leg{random} 
		
		\nextgroupplot[
		grid=none,
		tick label style={font=\large},
		tick align=inside,
		hide y axis,  
		axis x discontinuity=crunch,
		width=3.5cm, 
		height=5cm, 
		minor tick num=3,
		xmin=20, xmax=25,
		ymin=0, ymax=90,
		xtick={23.5},
		xticklabels={$2^{256}$},
		]

		\coordinate(a) at(23.5,0.0);
		\coordinate(b) at(23.5,0.438);
		\draw[-,dashed] (a) edge (b);
		
		\node[isosceles triangle, fill, \csdcolor, scale=0.4,rotate=30,isosceles triangle apex angle=60,] at (axis cs: 23.5, 83.05) {}; 
		\node[] at (axis cs:  22.9, 75.0) {PQ$_{32\|256}$}; 
		
	\end{groupplot}
	
\end{tikzpicture}

%% file: plots/dim_plot_par_medium_.tex
\pgfmathsetmacro{\teasermarkersize}{2.5}

\begin{tikzpicture}
	\begin{groupplot}[
		group style={
			group size=2 by 1,
			horizontal sep=0pt
		},
		axis x line=bottom,
		axis y line=middle, 
		longlines/.style={
			shorten >=-10mm,
			shorten <=-10mm
		},
		legend style={at={(axis cs:16.0, 25.0pt)}, anchor=south west, font=\normalsize, fill opacity=1.0},
		]
		
		\pgfplotstableread{
			gflops mixoxf csdoxf spectral
			10  27.7 21.2 27.35
			12  45.4 37.4 44.4
			14  49.7 46.1 49.45
			16  54.2 48.2 nan
		}{\map}
		
		\nextgroupplot[
		grid=none,
		height=5cm, 
		xlabel={Number of anchor points},
		ylabel={mAP ($\%$)},
		xmin=9, xmax=17,
		xtick={10,12,14,16},
		xticklabels={$2^{10}$,$2^{12}$, $2^{14}$, $2^{16}$},
		ymin=20, ymax=70,
		minor tick num=3,
		title={$\gR$Oxf and $\gR$Par (Hard)},
		title style={xshift=0.8cm, yshift=-5pt,font=\normalsize},
		grid style={line width=.1pt, draw=gray!5},
		major grid style={line width=.2pt,draw=gray!20},
		tick label style={font=\normalsize},
		tick align=inside,
		label style={font=\normalsize},
		legend style={font=\normalsize},
		x label style={xshift=0.8cm, yshift=0.1cm},
		y label style={yshift=-0.1cm},
		x axis line style={-},
		]
		
		\addplot[line width =2pt, oursoxf]      table[x=gflops,  y=mixoxf] \map; 
		\leg{\color{Maroon}{$k$-means}} 

            \addplot[line width =2pt, oursoxfs]      table[x=gflops,  y=spectral] \map; 
		\leg{\color{\ourscolors}{spectral}}
		
		\addplot[line width =2pt, thick, dotted, color=\csdcolor, mark=square]      table[x=gflops,  y=csdoxf] \map;
		\leg{random} 
		
		\nextgroupplot[
		grid=none,
		tick label style={font=\large},
		tick align=inside,
		hide y axis,  
		axis x discontinuity=crunch,
		width=3.5cm, 
		height=5cm, 
		minor tick num=3,
		xmin=20, xmax=25,
		ymin=20, ymax=70,
		xtick={23.5},
		xticklabels={$2^{256}$},
		]

		\coordinate(a) at(23.5,0.0);
		\coordinate(b) at(23.5,0.438);
		\draw[-,dashed] (a) edge (b);
		
		\node[isosceles triangle, fill, \csdcolor, scale=0.4,rotate=30,isosceles triangle apex angle=60,] at (axis cs: 23.5, 67.5) {}; 
		\node[] at (axis cs:  22.9, 63) {PQ$_{32\|256}$}; 
		
	\end{groupplot}
	
\end{tikzpicture}

%% file: plots/lightweight_query_models.tex
\resizebox{\linewidth}{!}{
\begin{tikzpicture}
	\begin{axis}[
		ybar=0pt,
		ymin=50,
		height=7cm,
		ymax=90,
		minor tick num=3,
		ylabel={mAP(\%)},
		every x tick label/.append style={font=\Large,rotate=15},
		every y tick label/.append style={font=\Large,rotate=0},
		title style={align=center, font=\Large, yshift=-5pt}, 
		title={$\mathcal{R}$Oxf and $\mathcal{R}$Par (Medium)},
		symbolic x coords={A, B, C, D, E},
		xtick=data,
		xticklabels={EB0, EB1, EB2, SV2, SV2$^{0.5 \times}$},
		xtick align=inside,
		ytick align=inside,
		ylabel near ticks,
		enlarge x limits=0.18,
		legend pos=north east,
		legend style={font=\large, align=right, row sep = -1pt, column sep=1.0pt,nodes={scale=1.0,transform shape}},
		legend columns=-1,
		grid=both,
		x tick label style={inner sep=1.0pt},
		legend image code/.code={\draw[#1, draw=none] (0cm,-0.2cm) rectangle (0.20cm,0.20cm);},
		grid style={line width=.1pt, draw=gray!5},
		major grid style={line width=.2pt,draw=gray!20},
		label style={font=\Large},
		]
		
		\addplot[ybar, fill=\ourscolor, bar width=15pt, postaction={pattern=north east lines,pattern color = lightgray}]
		coordinates {
			(A,  81.23)
			(B, 81.715)
			(C, 83.335)
			(D, 79.4 )
			(E, 73.61)
		};

		\addplot[ybar, fill=black, bar width=15pt, postaction={pattern=north west lines,pattern color = lightgray}]
		coordinates {
			(A,  75.445)
			(B,  77.54)
			(C, 79.59)
			(D, 66.58)
			(E, 58.895)
		};
		\legend{ Asysmmetric, Sysmmetric};
	\end{axis}
\end{tikzpicture}
\label{fig:bars_1_1}
\hspace{5pt}
\begin{tikzpicture}
	\begin{axis}[
		ybar=0pt,
		height=7cm,
		ymin=30,
		ymax=80,
		minor tick num=3,
		ylabel={mAP(\%)},
		every x tick label/.append style={font=\Large,rotate=15},
		every y tick label/.append style={font=\Large,rotate=0},
		title style={align=center, font=\Large, yshift=-5pt}, 
		title={$\mathcal{R}$Oxf and $\mathcal{R}$Par (Hard)},
		symbolic x coords={A, B, C, D, E},
		xtick=data,
		xticklabels={EB0, EB1, EB2, SV2, SV2$^{0.5 \times}$},
		xtick align=inside,
		ytick align=inside,
		ylabel near ticks,
		enlarge x limits=0.18,
		legend pos=north east,
		legend style={font=\large, align=right, row sep = -1pt, column sep=1.0pt,nodes={scale=1.0,transform shape}},
		legend columns=-1,
		grid=both,
		x tick label style={inner sep=1.0pt},
		legend image code/.code={\draw[#1, draw=none] (0cm,-0.2cm) rectangle (0.20cm,0.20cm);},
		grid style={line width=.1pt, draw=gray!5},
		major grid style={line width=.2pt,draw=gray!20},
		label style={font=\Large},
		]
		
		\addplot[ybar, fill=\ourscolor, bar width=15pt, postaction={pattern=north east lines,pattern color = lightgray}]
		coordinates {
			(A,  64.225)
			(B, 64.81)
			(C, 67.555)
			(D, 61.525)
			(E, 55.26)
		};

		\addplot[ybar, fill=black, bar width=15pt, postaction={pattern=north west lines,pattern color = lightgray}]
		coordinates {
			(A, 55.465)
			(B, 58.085)
			(C, 59.425)
			(D, 44.810)
			(E, 35.405)
		};
		\legend{ Asysmmetric, Sysmmetric};
	\end{axis}
\end{tikzpicture}
\label{fig:bars_1_2}
}

%% file: plots/PQ.tex
\begin{tikzpicture}[declare function={f(\x)=69.8;}]
\begin{axis}[%
	xmode=log,
	log ticks with fixed point,
  	height=7cm,
  	ymin=0,
  	ymax=80,
   	xlabel={Number of subspaces},
   	xtick={8,16,32,64,128,256},
   	ylabel={mAP(\%)},
  	minor tick num=3,
  	legend pos=south east,
        title={$\gR$Oxf + $\gR$1M and $\gR$Par + $\gR$1M (M)},
        title style={yshift=-5pt,font=\Large},
        grid style={line width=.1pt, draw=gray!5},
        major grid style={line width=.2pt,draw=gray!20},
        tick label style={font=\Large},
        label style={font=\LARGE},
        legend style={font=\large},
  ]

\pgfplotstableread{
nfeats  ResNet101 MobileNetv2
8      15.645  8.515 
16      21.535  15.45 
32      38.935  30.78 
64      56.505  50.445 
128     64.915  60.025 
256      68.20  63.80 
}{\map}

    \addplot[line width =2pt, thick=20pt, dotted, color=\csdcolor, mark=square]  table[x=nfeats,  y=ResNet101]   \map; 
    \leg{ResNet101}
    
    \addplot[oursoxf] table[x=nfeats,  y=MobileNetv2]   \map; 
    \leg{MobileNetV2}
    
\end{axis}
\end{tikzpicture}

{\begin{tikzpicture}[declare function={f(\x)=69.8;}]
		\begin{axis}[%
			height=7cm,
			xmode=log,
			ymin=0,
			ymax=50,
			log ticks with fixed point,
			xlabel={Number of subspaces},
			xtick={8,16,32,64,128,256},
			ylabel={mAP(\%)},
			minor tick num=3,
			legend pos=south east,
			title={$\gR$Oxf + $\gR$1M and $\gR$Par + $\gR$1M (H)},
			title style={yshift=-5pt,font=\Large},
			grid style={line width=.1pt, draw=gray!5},
			major grid style={line width=.2pt,draw=gray!20},
			tick label style={font=\Large},
			label style={font=\LARGE},
			legend style={font=\large, align=right, row sep = -1pt, column sep=1.0pt,nodes={scale=1.0,transform shape}},
			]
			
			\pgfplotstableread{
				nfeats  ResNet101 MobileNetv2
				8      4.295  3.01 
				16     6.625  4.385
				32     15.625 11.515 
				64     30.96  24.64
				128    40.01  35.295
				256    44.15  39.63
			}{\map}

			\addplot[line width =2pt, thick=20pt, dotted, color=\csdcolor, mark=square]  table[x=nfeats,  y=ResNet101]   \map; 
			\leg{ResNet101}
			
			\addplot[oursoxf] table[x=nfeats,  y=MobileNetv2]   \map; 
			\leg{MobileNetV2}
			
		\end{axis}
	\end{tikzpicture}}